\documentclass[pre,preprintnumbers,amsmath,amssymb,twocolumn]{revtex4-1}

\usepackage{graphicx,color}
\usepackage{bm}
\usepackage{epsfig}

\providecommand{\change}[1]{{#1}}

\newcommand{\mat}[1]{{\bf #1}}

\graphicspath{{data/}{figures/}}

\begin{document}

\title{Diffusion with stochastic resetting on a lattice}

\author{Alexander K. Hartmann}
\affiliation{Institut f\"ur Physik, Universit\"at Oldenburg, D-26111
             Oldenburg, Germany}
\author{Satya N. Majumdar}
\affiliation{LPTMS, CNRS, Univ.\ Paris-Sud,  Universit\'e Paris-Saclay,
  91405 Orsay,  France}

\begin{abstract}
We provide an exact formula for the mean first-passage time (MFPT) to a target at the origin for a single particle diffusing on 
a $d$-dimensional hypercubic {\em lattice} starting from a fixed initial position $\vec R_0$ and resetting
to $\vec R_0$ with a rate $r$. Previously known results in the continuous space are recovered in the
scaling limit $r\to 0$, $R_0=|\vec R_0|\to \infty$ with the product $\sqrt{r}\, R_0$ fixed.
However, our formula is valid for any $r$ and any
$\vec R_0$ that enables us to explore a much wider region of the parameter space that is inaccessible in the
continuum limit. For example, we have shown that the MFPT, as a function of $r$ for fixed $\vec R_0$, diverges 
in the two opposite limits
$r\to 0$ and $r\to \infty$ with a unique minimum in between, provided the starting point is not a nearest neighbour of
the target. In this case, the MFPT diverges as a power law $\sim r^{\phi}$ as $r\to \infty$,
but very interestingly with an exponent $\phi= (|m_1|+|m_2|+\ldots +|m_d|)-1$ that depends on the
starting point $\vec R_0= a\, (m_1,m_2,\ldots, m_d)$ where $a$ is the lattice spacing
and $m_i$'s are integers. If, on the other hand, the starting point
happens to be a nearest neighbour of the target, then the MFPT decreases monotonically with increasing $r$,
approaching a universal limiting value $1$ as $r\to \infty$, indicating that the optimal resetting rate in this
case is infinity. We provide a simple physical reason and a simple
Markov-chain explanation behind this somewhat unexpected universal result.
These interesting results on a lattice are not captured by the continuum theory. Our
analytical predictions are verified in numerical simulations on lattices up to $50$ dimensions.
Finally, in the absence of a target, we also compute exactly the position distribution of the walker
in the nonequlibrium stationary state that also displays interesting lattice effects not captured
by the continuum theory.

\end{abstract}


\maketitle

\date{\today}

\section{Introduction}

Stochastic resetting has emerged as an active area of research in statistical physics over the
past decade, finding numerous applications across diverse fields from stochastic processes to
random search algorithms. Stochastic resetting simply means interrupting the natural dynamics
of a process (classical or quantum) at random times and restarting the process. Perhaps
the simplest model
of stochastic resetting corresponds to a single particle diffusing in continuous space starting from  
a fixed initial position
and resetting to this initial position at a constant rate $r$~\cite{EM2011}. The exact solution of this model had
two interesting predictions~\cite{EM2011,EM2011b}: 
(i) the mean first-passage time (MFPT) to find a fixed target located at 
the origin is finite and as a function of $r$, displays a unique minimum at $r=r^*$ that depends
on the initial distance from the target and
(ii) in the absence of a target, the resetting at a constant rate $r$ drives the system
into a nonequilibrium stationary state at long times with a non-Gauusian position distribution.
Following (i),
one can then set the resetting rate at the minimum value $r^*$ to minimize the MFPT and thus expedite the search 
of the target. The existence of a finite optimal resetting rate $r^*$ has rendered the stochastc resetting 
as an efficient mechanism to speed up a search process.
This model subsequently triggered a flurry of activities where both aspects
of resetting (i) and (ii) were investigated, in theoretical models as well as in experiments (for reviews see,
\cite{EMS2020,PKR2022,GJ2022,GN2023}). For example, in one dimension, the MFPT of a diffusing particle 
starting at the initial position $x_0$ to a target located at the origin was found to
have a very simple expression~\cite{EM2011}
\begin{equation}
\langle T\rangle_r (x_0)= \frac{1}{r}\left[ e^{\sqrt{\frac{r}{D}}\, x_0}-1\right]\, .
\label{MFPT_d1_cont.1}
\end{equation}
The MFPT diverges in the two limits $r\to 0$ and $r\to \infty$ and has a minimum at some
intermediate $r^*$. The divergence as $r\to 0$ issues from the fact that 
in the absence of resetting the  `bad' trajectores
that take the walker away from the origin occur with high probability and contribute to the mean capture
time making it infinite. 
In the opposite limit, when there are many resettings, the trajectory essentially gets localised at
its starting position and the walker fails to reach the target making the MFPT divergent. This result was generalised to higher
dimensions~\cite{EM2014}, where the target needs to have a finite size in $d>2$ to be captured
with a nonzero probability by the random walker modelled as a point particle. 
These results for the MFPT in $d=1$ and $d=2$ were verified in
experiments on colloidal particles in optical traps~\cite{Roichman2020,Besga2020,Faisant2021}.

While the MFPT $\langle T\rangle_r(\vec R_0)$ of this single diffusing particle, starting
at the initial position $\vec R_0$ in $d$ dimensions and with a constant resetting rate $r$,
is well understood when the difusion takes place in continuous space, a natural question
is to wonder what happens to the MFPT for diffusion on a lattice in $d$ dimensions.
For example, does the MFPT on a lattice, as a function of $r$ and $\vec R_0$, exhibit
more or less similar behavior as in the continuous space, or does it have new interesting
regimes? The lattice computation is considerably more difficult than in the continuum, so
there is no point in just repeating the computations if they yield
qualitatively similar results.
But if there are new regimes as a function of $r$ and $\vec R_0$ in the lattice problem
that is not captured by the continuous space result, it would indeed be interesting
and worth pursuing the lattice calculations. 
\change{This question
was indeed addressed in Ref.~\cite{C2021} for a lattice random walker with resetting
in $d=1$ and some interesting differences from the continuum results
were observed. The purpose of this paper
is to solve the lattice model exactly in $d$-dimensions for arbitrary $d$ and
show that the lattice results display several interesting new features and also crucial differences
from their continuous counterparts in all dimensions. We will see that the
results from the lattice 
model are more general and includes the continuous space results in a special limit.
Our results are also verified by accompanying numerical
simulations and a simple Markov-chain representation.}

Some stochastic processes in the presence of resetting have been studied on lattices
~\cite{BS2014,MSS2015,FBGM2017,BFGM2019,BP2021,C2021,C2022,C2024,
BP2024} and networks~\cite{RBHM2020}. In Ref.~\cite{BMM2022},
a resetting random walk model on a $d$-dimensional lattice in discrete time
was studied, but the main objective there was to compute the mean number of distinct
sites visited by such a resetting random walker of $N$ steps.
\change{However, we have not
come across any explicit result for the MFPT for a random walker with resetting at rate $r$ on a
$d$-dimensional lattice for arbitrary $d$, except in the $d=1$ case in Ref.~\cite{C2021} mentioned
earlier.}

Let us briefly summarize our main results and also lay out the organization
of the paper. In Section \ref{Model}, we first define our model precisely on a lattice
and derive a general relation
between the MFPT in the presence of resetting and the free propagator of the underlying process without resetting.
This relation, based on renewal arguments, turns out to be very general and
holds for arbitrary Markov processes, both on the lattice
as well as in the continuum. This exact relation is given in Eq. (\ref{mfpt_ltg.1}).
In the next section \ref{lattice}, we evaluate this exact formula for a $d$-dimensional
hypercubic lattice leading to our main explicit formula in Eq. (\ref{mfpt_final}).
We show how to recover the continuous space results from this lattice MFPT formula,
in the limit when $r\to 0$, $R_0\to \infty$ but keeping $\sqrt{r}\, R_0$ fixed.

In Section \ref{sec:algorithm} we describe the event-driven algorithm we
have used to verify our analytical results numerically and to obtain results
on other properties of the walk, in particular the number of steps taken by the walker
to find the target.

In Section \ref{explicit}, we consider several special cases of our main formula
for lattice MFPT in Eq. (\ref{mfpt_final}) and derive explicit results.
We start with $d=1$ and $d=2$ and then present some exact results
in general dimensions in the two limits $r\to 0$ and $r\to \infty$ for a fixed starting position
$\vec R_0$. We denote the starting position $\vec R_0= a\, \vec m$ 
with  $\vec m\equiv (m_1,m_2,\ldots, m_d)$ where $m_i$'s are integers and $a$ is the lattice constant.
We show that as long as the starting position is not a nearest neighbour of the target
at the origin, the MFPT first decreases as a function of increasing $r$, achieves a global minimum
at some $r=r^*$ and then grows again and finally diverges as a power law as $r\to \infty$
\begin{equation}
\langle T\rangle_r(\vec R_0)\approx \frac{{\displaystyle {\prod_{i=1}^d }}\Gamma(|m_i|+1)}{
\Gamma\left({\displaystyle {\sum_{i=1}^d} |m_i| }+1\right)}\, r^{\phi}
\label{large_r_asymp.0}
\end{equation}
with  $\phi=(|m_1|+|m_2|+\ldots+|m_d|) -1$.
Thus the exponnent $\phi$ depends on the starting distance. 
\change{This generalizes a result in $d=1$ obtained in Ref.~\cite{C2021}}.
This interesting new regime
is inaccessible in the continuum theory where one has already taken the $r\to 0$ limit.
Another interesting explicit and general result is the following: we show that if the walker
instead starts from a lattice site which is an immediate neighbour of the target, then the
MFPT decreases monotonically with increasing $r$, achieving its minimum value $1$ (universal in
all dimensions) as $r\to \infty$, \change{thus generalizing
a similar observation made for $d=1$ in Ref.~\cite{C2021}}.
Thus in this case, the optimal resetting rate $r^*$ is actually infinite, while it is finite if the starting site
is not an immediate neighbour of the target! 
Hence, rather remarkably, for a starting site just next to the target, one needs to implement an
infinite reetting rate to minimize the MFPT! This is another strikingly surprising result that
is compleely missed by the continuum limit. We also compare our analytical predictions 
with numerical simulations up to $50$ dimensions, finding excellent agreement.
For some of our results we also provide a simple explanation based on
the properties of 
a few-state Markov process.

In the absence of a target, we show in Section \ref{NESS_section} that the system
is driven at long times into a nonequilibrium stationary state (NESS). We compute
the exact position distribution in the NESS for a resetting walker on a $d$-dimensional
hypercubic lattice.
Finally we conclude in Section \ref{Conclusion_section} and relegate some details of calculations
in Appendix \ref{Cont_lim_App}.

\section{Mean first-passage time to a target for a resetting random walker on a lattice: A general formula}
\label{Model}

We consider a resetting random walker on an infinite $d$-fimensional hypercubic lattice with lattice constant $a$. 
The position of the walker
evolves in continuous time,  starting at the initial position ${\vec R_0}$ at time $t=0$.
Let $\vec R$ be the current location of the walker at time $t$. Then in a small time $dt$,
the walker performs the following stochastic movements: 
\begin{itemize}

\item it hops to any one of the $2\, d$ neighbouring sites $\vec R +a\, \vec e$ of $\vec R$ with probability $dt$, i.e. with rate 
$\lambda_{\rm hop} = 1$

\item  it hops to the starting position $\vec R_0$ with probability $r\,dt$, where $r$ denotes the resetting rate

\item  with the remaining probability $1- (r+2d)\, dt$, it stays at the current location $\vec R$.

\end{itemize}

Let $P_r(\vec R, \vec R_0, t)$ denote the probability that the walker
is located  at site $\vec R$ at time $t$,
starting from $\vec R_0$ at $t=0$. The subscript $r$ in $P_r$ denotes the presence of 
resetting move with rate $r$.
This probability evolves via the 
Fokker-Planck equation 
\begin{widetext}
\begin{equation}
\frac{\partial P_r(\vec R,\vec R_0,\, t)}{\partial t}= \left[\sum_{\vec e} P_r(\vec R+a\, \vec e,\vec R_0,\, t) - 
2d\, P_r(\vec R,\vec R_0,\, t)\right] - r\, P_r (\vec R,\vec R_0,\, t) +r\, 
\delta_{\vec R, \vec R_0}\, ,
\label{fp.1}
\end{equation}
\end{widetext}
starting from the initial condition $P_r(\vec R, \vec R_0, 0)= \delta_{\vec R, \vec R_0}$ (Kronecker delta). 
The first term inside the parenthesis $[\cdots]$ on the
right hand side (rhs) of Eq. (\ref{fp.1}) is just a lattice Laplacian that
describes the diffusion, the second term describes the loss of probability from $\vec R$ due to resetting to $\vec R_0$
with rate $r$, while the third term describes the gain in probability at $\vec R_0$ via resetting from other sites.
We will later present an exact solution of the Fokker-Planck equation (\ref{fp.1}) in Section \ref{NESS_section}.

We note that Eq. (\ref{fp.1})  is just the lattice version of the Fokker-Planck equation in 
continuous space introduced and studied in Refs.~\cite{EM2011,EM2011b}. To see how the lattice Fokker-Planck
equation (\ref{fp.1}) reduces to its continuous counterpart, we take the continuum limit $a\to 0$, where $a$
is the lattice length. Expanding the right hand side (rhs) of Eq. (\ref{fp.1}) up to quadratic order in $a$, one gets
\begin{multline}
  \frac{\partial P_r(\vec R,\vec R_0,\, t)}{\partial t}\approx \\
  a^2\, \nabla_{\vec R}^2 P(\vec R, \vec R_0, t) - 
r\, P_r (\vec R,\vec R_0,\, t) +r\,
\delta_{\vec R, \vec R_0}\, ,
\label{fp_cont.1}
\end{multline}
Next we need to rescale time $t= \tilde{t}/a^2$ and consequently the resetting rate $r= \tilde{r}\, a^2$.
Furthermore, the probability density $\tilde{P}$ in continuous space is related to the probability
$P$ on a lattice via the relation $P=\tilde{P}\, a^d $.
Under this rescaling, Eq. (\ref{fp_cont.1}) reduces to the standard continuous Fokker-Planck equation studied
in Refs.~\cite{EM2011,EM2011b} (upon setting the diffusion constant $D=1$)
\begin{multline}
  \frac{\partial \tilde{P}_{\tilde{r}}(\vec R,\vec R_0,\, t)}{\partial \tilde{t}}= \\
  \nabla_{\vec R}^2 
\tilde{P}(\vec R, \vec R_0, t) - 
\tilde{r}\, \tilde{P}_{\tilde{r}}(\vec R,\vec R_0,\, t) +\tilde{r}\,
\delta\left(\vec R- \vec R_0\right)\, ,
\label{fp_cont.2}
\end{multline}
where we replaced $\delta_{\vec R, \vec R_0}/a^d$ by the Dirac delta function in the
limit $a\to 0$. Thus to recover the continuum limit, we
need to implement the following rescaling for the time and the resetting rate
\begin{equation}
\quad t= \frac{\tilde{t}}{a^2}\, \quad {\rm and} \quad r= \tilde{r}\, a^2\, ,
\label{rescaling.1}
\end{equation}
and take the limit $a\to 0$. Hence one
not only needs to take the large distance (in units of $a$) and late time limits, but needs additionally to take
the vanishing reset rate $r\to 0$ limit in an appropriate way.

We now assume that there is a target at the origin $\vec 0$ and we are interested in 
computing the MFPT $\langle T\rangle_r(\vec R_0)$ to the target, 
starting at $\vec R_0= a \,\vec m $, where $\vec m$ is a lattice site
$(m_1,m_2,\ldots, m_d)$  with
$m_i$'s integers. 
Let $F_r(\vec R_0, t)$ denote the first-passage probability density to the target, i.e,, the probability
density to reach the target for the first time at time $t$, starting at $\vec R_0$. 
Let $Q_r(\vec R_0,t)$ denote the survival probability of the walker up to time $t$, i.e., the probability
that it does not reach the target up to time $t$. Then clearly it is related to the first-passage probability
density via the simple relation~\cite{Redner_book,Persistence_review}
\begin{equation*}
  Q_r(\vec R_0,t)= \int_t^{\infty} F(\vec R_0, t')\, dt'
\end{equation*}
implying
\begin{equation}
F_r(\vec R_0,t)=- \frac{\partial Q_r(\vec R_0,t)}{\partial t}\, .
\label{rel.1}
\end{equation}
Consequently the MFPT can be expressed in terms of the survival probability
\begin{equation}
\langle T\rangle_r(\vec R_0)= \int_0^{\infty} t\, F_r(\vec R_0,t)\, dt= \int_0^{\infty} Q_r(\vec R_0, t)\, dt\, ,
\label{mfpt_surv.1}
\end{equation}
where we used the relaton in Eq. (\ref{rel.1}) and performed an integration by parts (assuming $Q_r(\vec R_0,t)$ decays
faster than $1/t$ as $t\to \infty $). Let us also define, for future usage, the Laplace transform of the survival
probability
\begin{equation}
\tilde{Q}_r(\vec R_0,s)= \int_0^{\infty} Q_r(\vec R_0, t)\, e^{-s\, t}\, dt\, .
\label{lap_def.1}
\end{equation}
Thus, the result (\ref{mfpt_surv.1}) then reduces to
\begin{equation}
\langle T\rangle_r(\vec R_0)= \tilde{Q}_r(\vec R_0, s=0)\, .
\label{mfpt_surv.2}
\end{equation}

Now, the survival probability $Q_r(\vec R_0,t)$ in the presence of resetting can be related to the survival
probability $Q_0(\vec R_0,t)$ in the absence of resetting, by using a renewal 
approach~\cite{EM2011,KMSS2014,Reuveni2016,EMS2020}. The renewal approach
leads to the exact equation
\begin{eqnarray}
  Q_r(\vec R_0,t) & = & e^{-r\, t}\, Q_0(\vec R_0,t) +   \\
 & & r\, \int_0^{t} d\tau\, e^{-r\tau}\,
  Q_0(\vec R_0,\tau)\, Q_r(\vec R_0, t-\tau)\, \nonumber
\label{renewal.1}
\end{eqnarray}
It is easy to understand the two terms on the rhs of Eq. (\ref{renewal.1}). The first term corresponds to the event
when there is no resetting in $[0,t]$ which happens with probability $e^{-r\, t}$. In this case, the survival
probability is simply $Q_0(\vec R_0,t)$ which, multiplied by the factor $e^{-r\, t}$, gives the first term.
The second term corresponds to the cases when there is one or more resetting events in $[0,t]$. In this case,
let $\tau$ denote the time at which the first resetting occurs. The
probability for the first resetting to occur in $[\tau, \tau+d\tau]$ is simply  $r\, e^{-r\tau}\, d\tau$.  
Prior to this resetting the target survives
with probability $Q_0(\vec R_0,\tau)$ and after the first resetting, the process renews and one needs to
multiply by $Q_r(\vec R_0, t-\tau)$ to ensure survival during the rest of the interval of duration $t-\tau$.
Finally the first resetting epoch $\tau$ can occur anywhere in $[0,t]$ and one needs to integrate over $\tau\in [0,t]$
to obtain the second term in \eqref{renewal.1}. The convolution structure of the second term in \eqref{renewal.1}
immediately
suggests to take a Laplace transform of this equation with respect to $t$. Performing this Laplace transform,
one gets a very simple general relation
\begin{equation}
\tilde{Q}_r(\vec R_0,s) = \frac{\tilde{Q}_0(\vec R_0, s+r)}{1-r\, \tilde{Q}_0(\vec R_0, s+r)}\, .
\label{rel.3}
\end{equation}
Let us note that there is a shift in the Laplace variable from $s$ to $s+r$ on the rhs of \eqref{rel.3}.
Consequently setting $s=0$ and using \eqref{mfpt_surv.2}, the MFPT in the presence of resetting can be
expressed explicitly in terms of the Laplace transform of the survival probability in the absence
of resetting
\begin{equation}
\langle T\rangle_r(\vec R_0)= \frac{\tilde{Q}_0(\vec R_0, r)}{1-r\, \tilde{Q}_0(\vec R_0, r)}\, .  
\label{mfpt_surv.3}
\end{equation}
It is convenient to express the rhs of Eq. (\ref{mfpt_surv.3}) in terms of the Laplace transform of
the first-passage probability density (without resetting). This can be done by recallng that
that the first-passage probability density is related to the survival probabilty via the relation
\begin{equation}
F_0(\vec R_0,t)= -\partial_t Q_0(\vec R_0,t) \, .
\label{fp_surv.0}
\end{equation}
Taking Laplace transform with respect to $t$ yields
\begin{equation}
\tilde{F}_0(\vec R_0,s)=1- s\, \tilde{Q}_0(\vec R_0, s)\, .
\label{survival_fp.1}
\end{equation}
Using this relation in Eq. (\ref{mfpt_surv.3}), the MFPT can then be expressed as
\begin{equation}
\langle T\rangle_r(\vec R_0)= \frac{1}{r}\left[ \frac{1}{\tilde{F}_0(\vec R_0,r)}-1\right]\, .
\label{mfpt_fp.1}
\end{equation}
Thus, thanks to this explicit relation, computing the MFPT in the presence of resetting just requires
the knowledge of the Laplace transform of the first-passage probability density in the absence of resetting. 

We now show how to express
this Laplace transform $\tilde{F}_0(\vec R_0, s)$ for a pure (without resetting)
random walker on a $d$-dimensional lattice in terms of the lattice Green's function.
To find this relation,
consider a random walk trajectory that starts at $\vec R_0$ and arrives at the origin at time $t$. The
probability for this event is simply $P_0(\vec 0, \vec R_0, t)$. 
Now, this path that arrives at the origin at time $t$
must have hit the origin for the first time at some epoch before $t$,
say at $\tau$ 
and then has
returned to the origin at time $t$
(possibly $t=\tau$ if it is the first visit). Hence one can again use a renewal equation to connect the
first-passage probability density and the return probability
\begin{equation}
P_0(\vec 0, \vec R_0, t) = \int_0^{t} d\tau F_0(\vec R_0, \tau)\, P_0(\vec 0, \vec 0, t-\tau)\, .
\label{renewal.2}
\end{equation}
Taking Laplace transform with respect to time gives
\begin{equation}
\tilde{F}_0(\vec R_0,s)= \frac{\tilde{P}_0(\vec 0, \vec R_0,s)}{\tilde{P}_0(\vec 0, \vec 0, s)}  
\quad {\rm for}\quad \vec R_0\ne \vec 0\, .
\label{fp_ltg.1}
\end{equation}
The Laplace transform $\tilde{P}_0(\vec R, \vec R_0, s)$ ( or its analogue generating function
when the walk takes place in discrete time steps) is usually referred to as the lattice Green's 
function (for a nice review on lattice Green's function, see Ref.~\cite{Guttmann2010}). 
The ratio of these lattice Green's function also appears in the computation
of the mean number of distinct sites visited by a random walker on a lattice~\cite{Vineyard1963,MW1965},
see also the recent Ref.~\cite{BMM2022} in the context of a resetting random walker. 

Substituting the result \eqref{fp_ltg.1} in Eq. (\ref{mfpt_fp.1}) then expresses the MFPT with resetting
in terms of the lattice Green's function without resetting
\begin{equation}
\langle T\rangle_r(\vec R_0)=\frac{1}{r}\left[ \frac{\tilde{P}_0(\vec 0, \vec 0, r)}{\tilde{P}_0(\vec 0, 
\vec R_0, r)}-1\right]\, .
 \label{mfpt_ltg.1}
\end{equation}
We note that this result (\ref{mfpt_ltg.1}) is actually very general, and holds even for random walks
in the presence of a force or potential. It is valid in all dimensions. It is also
valid for a Brownian motion in $d$-dimensions in the presence or absence of a drift, where
$\tilde{P}_0(\vec R, \vec R_0,s)$ is just the Laplace transform of the propagator of the process.
For example, for a Brownian motion in one dimension, the propagator in real time is simply
\begin{equation}
P_0(R, R_0,t) = \frac{1}{\sqrt{4\,\pi\, D\, t}}\, e^{- (R-R_0)^2/{4Dt}}\, .
\label{bm.1}
\end{equation}
Its Laplace transform is simply
\begin{eqnarray}
\tilde{P}_0(\vec R, \vec R_0, s) & = & \int_0^{\infty} dt\, e^{-s\, t}\, 
\frac{1}{\sqrt{4\,\pi\, D\, t}}\, e^{- (R-R_0)^2/{4Dt}} \nonumber \\
& =&\frac{1}{\sqrt{4\, s, D}}\, e^{- \sqrt{s/D}\, |R-R_0|}\, .
\label{bm.2}
\end{eqnarray}
Substituting (\ref{bm.2}) in \eqref{mfpt_ltg.1}, one recovers the continuous space result 
in Eq. (\ref{MFPT_d1_cont.1}).

The expression in Eq. (\ref{mfpt_ltg.1}) is our main result which says that
to compute the MFPT in the presence of resetting, we just need
to know the Laplace transform of the propagator of the underlying process without resetting.
We show in the next subsection
how to evaluate it explicitly for a random walk on a $d$-dimensional hypercubic lattice.

\section{An explicit formula for the MFPT for a resetting $d$-dimensional random walk
on a hypercubic lattice}
\label{lattice}

Consider a reset free ($r=0$) random walk, whose position distribution evolves via
the lattice diffusion equation
\begin{multline}
  \frac{\partial P_0(\vec R, \vec R_0, t)}{\partial t}= \\
  \left[\sum_{\vec e} P_0(\vec R+a\, \vec e,\vec R_0, t)- 2d\, 
P_0(\vec R, \vec R_0, t)\right]\, , 
\label{fp_r0.1}
\end{multline}
obtained by setting $r=0$ in Eq. (\ref{fp.1}). It starts from the initial condition 
$P_0(\vec R, \vec R_0,\, t=0)= \delta_{\vec R, \vec R_0}$. This linear equation can be solved exactly using the
Laplace-Fourrier transform~\cite{MW1965}. 
Let us first take the Laplace transform of (\ref{fp_r0.1})
with respect to $t$, by defining
\begin{equation}
\tilde{P}_0(\vec R, \vec R_0, s)= \int_0^{\infty} P_0(\vec R, \vec R_0, t)\, e^{-s\, t}\, dt\, .
\label{lap_def.2}
\end{equation}
Using the initial condition, we get
\begin{equation}
(s+2d)\, \tilde{P}_0(\vec R, \vec R_0, s)-\delta_{\vec R, \vec R_0}= 
{\displaystyle \sum_{\vec e}} \tilde{P}_0(\vec R+a\,\vec e, \vec R_0, s)\, .
\label{fp_lap.1}
\end{equation}

Next we define the Fourier transform
\begin{equation}
{\hat P}_0(\vec k, s)= {\displaystyle \sum_{\vec R}} \tilde{P}_0(\vec R, \vec R_0, s)\, 
e^{i\, \vec k  \cdot \frac{(\vec R-\vec R_0)}{a}}\, .
\label{fp_Fourier.1}
\end{equation}
Taking Fourier transform of \eqref{fp_lap.1} gives
\begin{equation}
{\hat P}_0(\vec k, s)= \frac{1}{\left[(s+2d) - 2\, {\displaystyle \sum_{i=1}^d } \cos(k_i)\right]}\, .
\label{fp_Fourier.2}
\end{equation}
Finally, inverting the Fourier transform and using symmetries, one gets the well known expression for the
lattice Green's function~\cite{MW1965,Guttmann2010}
\begin{multline}
  \tilde{P}_0(\vec R, \vec R_0, s)= \\
  \int_{-\pi}^{\pi} \frac{dk_1}{2\pi}\ldots \int_{-\pi}^{\pi} \frac{dk_d}{2\pi}
\frac{e^{-i\, \vec k\cdot \frac{(\vec R-\vec R_0)}{a} }}{\left[(s+2d) - 2\, {\displaystyle \sum_{i=1}^d } \cos(k_i)\right]}\, .
\label{lattice_gf.1}
\end{multline}
We next set $\vec R=\vec 0$ and $\vec R_0= a\, \vec{m}$ where
$\vec m\equiv (m_1,m_2,\ldots, m_d)$ with $m_i$'s being 
integers. 
Using the symmetry in the
$k$ space,  we can then express Eq. (\ref{lattice_gf.1})
as a $d$-dimensional real integral
\begin{multline}
  \tilde{P}_0(\vec 0, \vec R_0=a\, {\vec m}, s)= \\
  \int_{-\pi}^{\pi} \frac{dk_1}{2\pi}\ldots \int_{-\pi}^{\pi} \frac{dk_d}{2\pi}
\frac{ {\displaystyle \prod_{i=1}^d} \cos(k_i\, m_i) }{\left[(s+2d) - 2\, {\displaystyle \sum_{i=1}^d } \cos(k_i)\right]}\, .
\label{lattice_gf.2}
\end{multline}

To make further progress, we use the integral representation
\begin{equation}
\frac{1}{z}= \int_0^{\infty} dt\, e^{-t\, z}\, ,
\label{int_rep.1}
\end{equation}
to rewrite Eq. (\ref{lattice_gf.2}) as
\begin{multline}
  \tilde{P}_0(\vec 0, \vec R_0=a\, {\vec m}, s)= \frac{1}{(s+2d)} \times \\
 \int_0^{\infty} dt\, e^{-t}\, 
{\displaystyle \prod_{i=1}^d} \int_{-\pi}^{\pi} \frac{dk_i}{2\pi}\, 
e^{- \frac{2\,t}{(s+2d)}\, \cos(k_i)}
\, \cos(k_i\, m_i)\, .
\label{int_rep.2}
\end{multline}
Next we use the identity~\cite{Grads}
\begin{equation}
\int_{-\pi}^{\pi} \frac{dk}{2\pi}\, \exp\left[ z\, \cos(k)\right] \, \cos(k\, m)= I_{|m|}(z)\, ,
\label{iden.1}
\end{equation}
where $I_m(z)$ is the modified Bessel function of the first kind with index $m$ and argument $z$. 
This identity \eqref{iden.1} is valid
only when $m$ is an integer, as in our case. Using \eqref{iden.1} in (\ref{int_rep.2}) we get a compact formula
\begin{multline}
  \tilde{P}_0(\vec 0, \vec R_0=a\, {\vec m}, s)=\\
  \frac{1}{(s+2d)}\int_0^{\infty} dt\, e^{-t}\,
{\displaystyle \prod_{i=1}^d} I_{|m_i|}\left(\frac{2\, t}{(s+2d)}\right)\, .
\label{int_rep.3}
\end{multline}
Substituting this result in Eq. (\ref{mfpt_ltg.1}), we get an explicit expression for the MFPT of
the lattice walker with resetting rate $r$ and starting from
$\vec R_0=a\, (m_1,m_2,\dots, m_d)$
\begin{equation}
\langle T\rangle_r\left({\vec R_0}=a\, {\vec m}\right)=\frac{1}{r}\left[ \frac{ \int_0^{\infty} dt\, e^{-t}\,
\left[I_{0}\left(\frac{2\, t}{(r+2d)}\right)\right]^d}{\int_0^{\infty} dt\, e^{-t}\,
{\displaystyle \prod_{i=1}^d} I_{|m_i|}\left(\frac{2\, t}{(r+2d)}\right)}-1\right]\, .
\label{mfpt_final}
\end{equation}
This is our main new general result valid in arbitrary dimension and arbitrary starting point
$\vec R_0=a\, (m_1,m_2,\ldots, m_d)$. Unfortunately, the integrals on the rhs of
\eqref{mfpt_final} can not be computed in closed form for general
dimension $d$ and general $\vec R_0$.
However, the representation in \eqref{mfpt_final} is still nice because it can be
easily evaluated numerically in Mathematica, for any $d$ and any choice of $\vec R_0$.
We analyse later \eqref{mfpt_final} in several special cases where the integrals can be done
exactly to make the formula for the MFPT even more explicit.

\subsection{Recovering the continuous space limit results} 

To recover the continuous space results from
our exact lattice MFPT in Eq. (\ref{mfpt_final}), we need to take $a\to 0$ limit and also rescale
the resetting rate $r= a^2\, \tilde{r}$ and define the rescaled MFPT $\tilde{T}= a^2\, T$ (see Eq. (\ref{rescaling.1})).
In Appendix \ref{Cont_lim_App} we show in detail how to take this limit in Eq. (\ref{mfpt_final}) to obtain, in arbitrary dimenion $d$,
\begin{multline}
  \langle \tilde{T}\rangle_{\tilde{r}}({\vec R_0})= \lim_{a\to 0} \langle T\rangle_{r=a^2\, \tilde{r}}({\vec R_0}) = \\
  \frac{1}{\tilde{r}}\, \left[ \frac{\Gamma(|\nu|)\, ^{|\nu|-1}\, (\tilde{r})^{-|\nu|/2}\, a^{\nu-|\nu|}}{
R_0^{\nu}\, K_{\nu}\left(R_0\, \sqrt{\tilde{r}}\right)}-1\right]
\label{result_cont.1}
\end{multline}
with $\nu= 1-\frac{d}{2}$ and
  where $K_\nu(z)$ is the modified Bessel function of the second kind with index $\nu$ and argument $z$.
For $d=2$, i.e., $\nu=0$, one obtains a 
slightly different behavior with a logaritmic correction. For $d<2$, the limit $a\to 0$ exists and is finite, while
for $d>2$, one needs to keep a small but nonzero lattice constant $a$. In Ref.~\cite{EM2014}, the MFPT was
computed directly in the continuum limit by assuming that the target is spherical with a finite radius $\epsilon$
and it was found that
\begin{equation}
\langle \tilde{T}\rangle_{\tilde{r}}({\vec R_0})= 
\frac{1}{\tilde{r}}\, \left[ \frac{\epsilon^{\nu}\, K_{\nu}\left(\epsilon\, \sqrt{\tilde{r}}\right)}{
R_0^{\nu}\, K_{\nu}\left(R_0\, \sqrt{\tilde{r}}\right)}-1\right]\, .
\label{result_cont.2}
\end{equation}
One can now take the limit $\epsilon\to 0$ using the following asymptotic small $z$ behavior of the Besssel function
\begin{eqnarray}
K_\nu(z) \approx  \begin{cases}
& \Gamma(|\nu|)\, 2^{|\nu|-1}\, z^{-|\nu|}\quad {\rm as}\, z\to 0\,, |\nu|>0 \\
\\
& -\ln (z/2)- \gamma_E \, \quad {\rm as} \, z\to 0\,, |\nu|=0\, ,
\end{cases}
\label{BesselK_asymp}
\end{eqnarray}
where $\gamma_E$ is the Euler gamma constant. Using this in Eq. (\ref{result_cont.2}), one indeed recovers
Eq. (\ref{result_cont.1}) upon identifying $\epsilon$ with the lattice constant $a$.
Finally, let us remark again that the continuum result \eqref{result_cont.1} is less richer
than our lattice result in \eqref{mfpt_final}. The lattice MFPT result \eqref{mfpt_final} has some 
interesting new regimes, 
in particular for large resetting rate $r$ with fixed
starting point $\vec R_0$, that the continuum limit misses since one needs to already take the limit 
$r\to 0$ in arriving
at the continuum limit.

\section{Numerical approach \label{sec:algorithm}}

We implemented a simple continuous-time event-driven algorithm to simulate
the behavior of the resetting random walker and to compare
the numerical results
to our analytical predictions. This algorithm also enables us to measure
other quantities of interest, e.g., the total number of hops till the target is found
and the number of hops needed to capture the target after the final reset.

The main idea of the event-based approach is to generate the times of the
relevant events, hops and resets, and to perform changes to the walker's
position only at event times. Events that take place with some rate $\lambda$
follow an exponential distribution with parameter $\lambda$, given by
the probability density $p(t)=\lambda e^{-\lambda t}$ $(t>0)$,
i.e., the probability distribution $P(t)=\int_0^t p(t')\, dt'=1-e^{-\lambda t}$.
Actually we use $\Delta t$ to denote the duration to the next event instead
of $t$, which denotes the total time here.
Since the distribution function can be inverted, the duration
 until the next event
can be simply generated \cite{practical_guide2015}
by using the \emph{Inversion Approach}. This means, one draws
a random number
$u$ which is uniformly distributed  in the interval $[0,1]$.
Then one applies  the inverse distribution,
i.e., assigns $\Delta t = P^{-1}(u)=\log(1-u)/\lambda$. One writes
$\Delta t \sim $Exp($\lambda)$ to indicate this generation.

Within the simulations, we assume lattice constant $a=1$.
We denote by $\vec R$ the current
position of the walker, by $t$ the time of the last event, by $t_r$ the
time of the next reset event and by $t_{\rm hop}$ the time of the
next hopping event.
Each walker is considered until the target is found.

A hopping event will happen with rate $2d$ since
each of the possible $2d$ hop occurs with
rate $\lambda_{\rm hop}=1$. Since all hops to the
neighbors have this same rate, for each occurring hop,
one of the $2d$ directions will be
chosen randomly with probability $1/(2d)$. We also measure the total number
$n_{\rm hop}$ of hops and the number $n_{\rm final}$ of hops since
the last reset.
The algorithm reads as follows:

\begin{tabbing} xx \= xx \= xx \= xx \= xx \= xx \= xx \= xxxxxxxxxxxxx \= \kill
{\bf algorithm} walk($\vec R_0$, $r$, $d$) \\
{\bf begin}\\
\> $\vec R = \vec R_0$; \>\>\>\>\>\>\> $\setminus\setminus$ initialise \\
\> $t=0$;\\
\> $n_{\rm hops} =0$; $n_{\rm final}=0$ \\
\> $ \Delta t \sim $Exp$(r)$; $t_r=t+\Delta t$;
\>\>\>\>\>\>\> $\setminus\setminus$ first reset  \\
\> $ \Delta t \sim $Exp$(2d)$; $t_{\rm hop}=t+\Delta t$;
\>\>\>\>\>\>\> $\setminus\setminus$ first hop \\
\> {\bf while} $\vec R \neq \vec 0$  {\bf do}
                 \>\>\>\>\>\>\> $\setminus\setminus$ while $\urcorner$found\\ 
\> {\bf begin}\\
\>\> {\bf if } $t_r < t_{\rm hop}$ {\bf then}\\
\>\> {\bf begin} \>\>\>\>\>\> $\setminus\setminus$ perform reset\\
\>\>\> $\vec R = \vec R_0$;\\
\>\>\> $n_{\rm final}= 0$;\\
\>\>\> $t=t_r$;\\
\>\>\> $ \Delta t \sim $Exp$(r)$; $t_r=t+\Delta t$;
\>\>\>\>\> $\setminus\setminus$ next reset \\
\>\> {\bf end}\\
\>\> {\bf else}\\
\>\> {\bf begin} \>\>\>\>\>\> $\setminus\setminus$ perform hop\\
\>\>\> random direction $\vec d=\pm \vec e_i$;\\
\>\>\> $\vec R = \vec R + \vec d$;\\
\>\>\> $n_{\rm hop}= n_{\rm hop}+1$;\\
\>\>\> $n_{\rm final}= n_{\rm final}+1$;\\
\>\>\> $t=t_{\rm hop}$;\\
\>\>\> $ \Delta t \sim $Exp$(2d)$; $t_{\rm hop}=t+\Delta t$;
\>\>\>\>\> $\setminus\setminus$ next hop\\
\>\> {\bf end}\\
\> {\bf end}\\
\> {\bf return}($t$, $n_{\rm hops}$, $n_{\rm final}$)\\
{\bf end}\\
\end{tabbing}

Note that all event rates are independent of the lattice sites,
so after a hop or reset, no other events need to be rescheduled.

\section{Some explicit results}
\label{explicit}

Our starting point is the exact lattice MFPT formula in Eq. (\ref{mfpt_final}).
We will now analyse this formula explicitly in several special cases.
For convenience, in the rest of the paper, we will use the following notation
\begin{equation}
\langle T\rangle_r(\vec R_0= a\, \vec m)\equiv \langle T\rangle_r(m_1,m_2,\ldots, m_d)\, .
\label{MFPT_newdef}
\end{equation}

\subsection{Explicit results in $d=1$}

\begin{figure}
\includegraphics[width=0.45\textwidth]{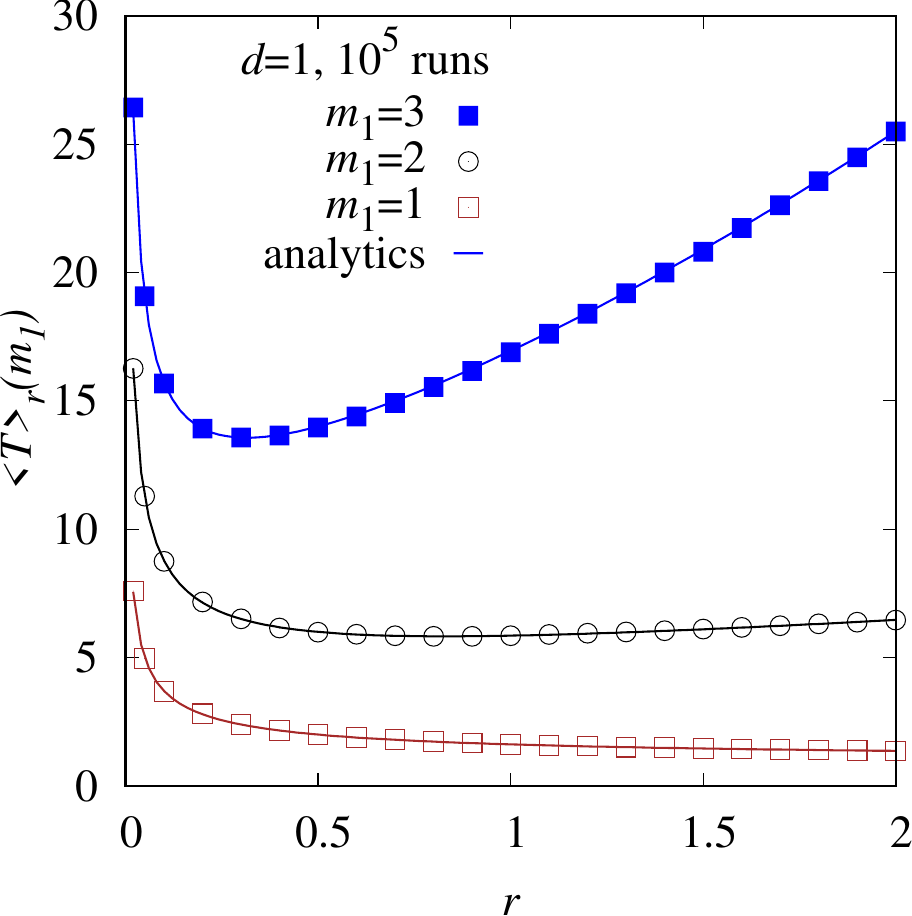}
\caption{$\langle T\rangle_r(m_1)$ vs. $r$ in one dimension ($d=1$) for $m_1=1$, $m_1=2$ and $m_1=3$. While for any $m_1>1$, the
curve exhibits a unique minimum at some $r=r^*(|m|)$, for $m_1=1$ it decays monotonically with increasing $r$,
approaching the limiting value $1$ as $r\to \infty$.
The solid lines correspond to the analytical formula
in Eq.\ (\ref{mfpt_d1.1}). } 
\label{fig_d1}
\end{figure}

In $d=1$, the result for the MFPT becomes fully explicit via the identity~\cite{Grads}
\begin{equation}
\int_0^{\infty} dt\, e^{-t}\, I_{|m|}(z\, t)= \frac{z^{-|m|}\left[1-\sqrt{1-z^2}\right]^{|m|}}{\sqrt{1-z^2}}\, .
\label{iden.2}
\end{equation}
Using this identity in Eq. (\ref{mfpt_final}) for $d=1$ and simplifying, we get the explicit result
\begin{equation}
\langle T\rangle_r(m_1)= \frac{1}{r}\left[ \left(\frac{2}{r+2-\sqrt{r^2+4\, r}}\right)^{|m_1|}-1\right]\, ,
\label{mfpt_d1.1}
\end{equation}
valid for any integer $m_1$. \change{We note that this result
was in $d=1$ was already derived in Ref.~\cite{C2021} by a different method}.
As a function of $r$ for fixed $|m_1|$, it has the asymptotic behaviors
\begin{eqnarray}
\label{mfpt_asymp}
\langle T\rangle_r(m_1)\approx 
\begin{cases}
& \frac{|m_1|}{\sqrt{r}} \quad\quad\, {\rm as}\quad r\to 0  \\ 
\\
& r^{|m_1|-1} \quad {\rm as} \quad r\to \infty \, .
\end{cases}
\end{eqnarray}
Thus, as $r\to 0$, it diverges for any $|m_1|$. In contrast, as $r\to \infty$, it diverges as a power law for any $|m_1|>1$,
but it aproaches a constant $1$ for $|m_1|=1$, \change{thus reproducing the observations made in Ref.~\cite{C2021} for $d=1$.} 
In Fig.~\ref{fig_d1}, we plot the MFPT $\langle T\rangle_r(m_1)$ in Eq. (\ref{mfpt_d1.1})
as a function of
$r$ for three different values of $m_1=1,\, 2,\, 3$ (shown by solid lines) and compare
them to direct numerical simulation results (symbols), finding perfect matching. 
For any $|m_1|>1$, it then displays a unique
minimum at some $r=r^*(|m|)$. However, for $|m_1|=1$, it decreases monotonically to its asymptotic value
$1$ as $r\to \infty$. We will see later that this is a generic feature in any dimension.
If the walker starts from a site that is nearest neighbour to the origin (target), then the
MFPT decreases monotonically as $r$ increases, approaching the universal value $1$ in all
dimensions as $r\to \infty$. However, for any starting site that is not a nearest neighbour
of the origin, the MFPT displays a unique minimum at some $r^*$ that depends on the starting site and
the dimension $d$. We will provide later a simple physical argument for this universal
asymptotic value $1$ of the MFPT as $r\to \infty$ when the starting site is a nearest neighbour of
the origin.

Finally, let us note that by setting $m_1=R_0/a$, $T= \tilde{T}\, a^2$, $r=\tilde{r} a^2$ in
Eq. (\ref{mfpt_d1.1}) and taking the $a\to 0$ limit, we get
\begin{equation}
\langle \tilde{T}_{\tilde{r}}(\vec R_0)= \frac{1}{\tilde{r}}\left[e^{\sqrt{\tilde{r}}\, R_0}-1\right]\, ,
\label{d1_cont.1}
\end{equation}
thus recovering the continuous space result in Eq. (\ref{MFPT_d1_cont.1}) with diffusion constant $D=1$ set to unity.
Thus the continuum limit corresponds to taking the limits $r\to 0$, $|m_1|\to \infty$ keeping
the product $\sqrt{r}\, |m_1|$ fixed in Eq. (\ref{mfpt_d1.1}). Consequently, in the $(r,m_1)$ plane,
the continuum limit captures the behaviour of the MFPT only in one corner. However, the lattice result
\eqref{mfpt_d1.1} holds in the full $(r, m_1)$ plane and hence is richer. In particular,
the power law growth of the MFPT $r^{|m_1|-1}$ for large $r$ in Eq. (\ref{mfpt_asymp} is
completely missed by the continuum limit.

\subsection{Explicit results in $d=2$}

In $d=2$, the MFPT in Eq. (\ref{mfpt_final}) reads for the starting
point $\vec R_0=a\, (m_1,m_2)$
\begin{multline}
  \langle T\rangle_r(m_1,\, m_2)=\\
  \frac{1}{r}\left[ \frac{ \int_0^{\infty} dt\, e^{-t}\,
\left[I_{0}\left(\frac{2\, t}{(r+4)}\right)\right]^2}{\int_0^{\infty} dt\, e^{-t}\,
I_{|m_1|}\left(\frac{2\, t}{(r+4)}\right)\, I_{|m_2|}\left(\frac{2\, t}{(r+4)}\right)}   -1\right]\, .
\label{mfpt_d2.1}
\end{multline}
The integral in the numerater on the rhs of Eq. (\ref{mfpt_d2.1}) can be computed 
explicitly for $|z|<1$~\cite{Guttmann2010}
\begin{equation}
\int_0^{\infty} dt\, e^{-t}\,\left[I_{0}\left(\frac{z\, t}{2}\right)\right]^2= \frac{2}{\pi}\,
K(z^2)= {}_2F_1\left(\frac{1}{2}, \frac{1}{2}, 1, 
z^2\right)\, ,
\label{num_2d.1}
\end{equation}
where $K(u)$ is the EllipticK 
function with argument $u$ and  ${}_2F_1$ is the standard hypergeometric function~\cite{Grads}.
The EllipticK function is defined as
\begin{equation}
K(u)= \int_0^{\pi/2} \frac{d\theta}{\sqrt{1- u\, \sin^2(\theta)}}\, .
\label{K_func_def}
\end{equation}
Note that there was a typographical error 
in Ref.~\cite{Guttmann2010}
where the argument of the hypergeometric function was reported to be $z$, instead of $z^2$.
In contrast, the integral in the denominator on the rhs of Eq. (\ref{mfpt_d2.1}) does not seem to be doable
for general $(m_1, m_2)$. However, it can be done explicitly for the two cases $(1,0)$ and $(1,1)$.

\vskip 0.3cm

\begin{figure}
\includegraphics[width=0.45\textwidth]{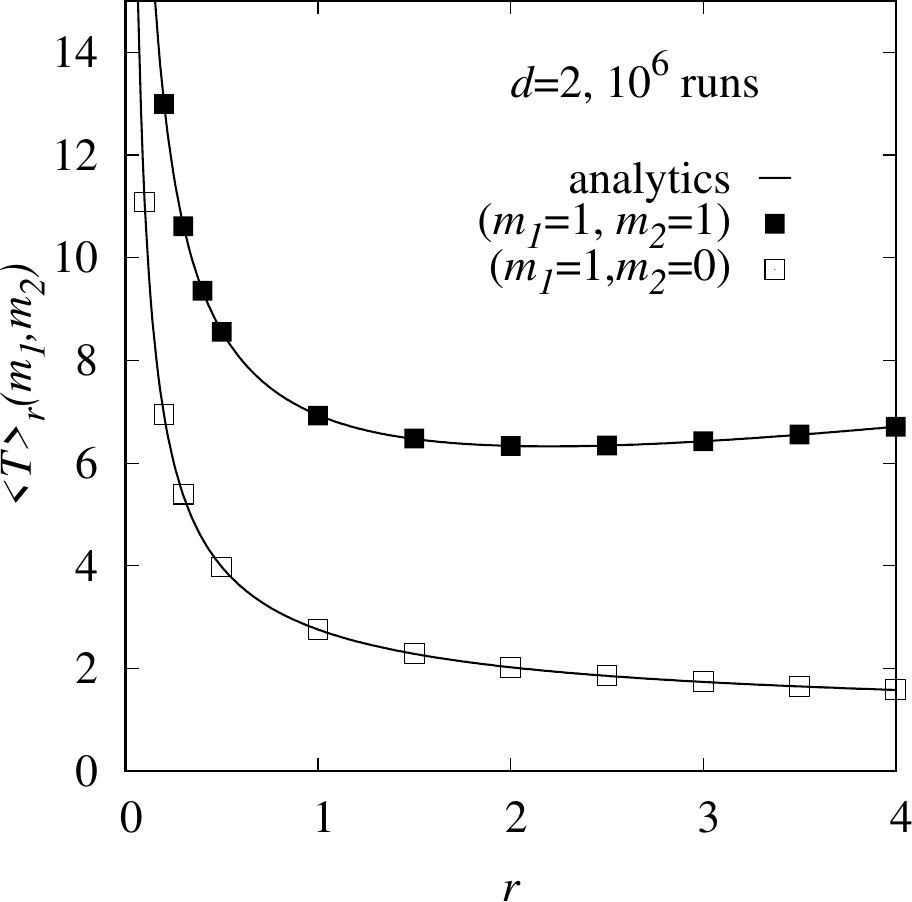}
\caption{$\langle T\rangle_r(1,0)$ and $\langle T\rangle_r(1,1)$ vs. $r$
  in two dimensions.
  The solid line shows the analytical results in Eqs.\ (\ref{mfpt_d210.2})
  and (\ref{mfpt_d211.2}),
  while the symbols represent
the numerical results. The agreement is excellent. 
}
\label{fig_d2mfpt1x}
\end{figure}

\noindent {\bf The case $(m_1=1,m_2=0)$.} In this case, Eq. (\ref{mfpt_d2.1}) reduces to
\begin{equation}
\langle T\rangle_r(1,\, 0)= \frac{1}{r}\left[\frac{2}{\pi}\, \frac{K\left(\frac{16}{(4+r)^2}\right)}{\int_0^{\infty} dt\, e^{-t}\,
I_{0}\left(\frac{2\, t}{(r+4)}\right)\, I_{1}\left(\frac{2\, t}{(r+4)}\right)}-1\right]\, .
\label{mfpt_d210.1}
\end{equation}
To compute the integral in the denominator in \eqref{mfpt_d210.1} we first use
the identity $I_1(z)=I_0'(z)$~\cite{Grads} to re-write this integral as
\begin{equation}
\int_0^{\infty} dt\, e^{-t}\, I_0(zt/2)\, I_1(zt/2)= \frac{1}{z}\, \int_0^{\infty}
dt\, e^{-t}\, \frac{d}{dt}[ I_0^2(zt/2)]\, .
\label{iden.3}
\end{equation}
Performing the last integral by parts and using the result from \eqref{num_2d.1} we get
\begin{equation}
\int_0^{\infty} dt\, e^{-t}\, I_0(zt/2)\, I_1(zt/2)= \frac{1}{z}\left[\frac{2}{\pi} K(z^2)-1\right]\, ,
\label{iden.4}
\end{equation}
valid again for $|z|<1$.
Substituting in \eqref{mfpt_d210.1} and simplifying leads to the exact result
\begin{equation}
\langle T\rangle_r(1,\, 0)= \frac{1}{r(r+4)}\left[\frac{4}{\frac{2}{\pi}\, K\left(\frac{16}{(4+r)^2}\right)-1}-r\right]\, .
\label{mfpt_d210.2}
\end{equation}
The asymptotic behaviors are given by
\begin{eqnarray}
\label{10_asymp}
\langle T\rangle_r(1,\, 0)\approx \begin{cases}
& -\frac{\pi}{r\, \ln (r)} \quad {\rm as} \quad r\to 0 \\
\\
& 1+ \frac{3}{r} \quad\quad\, {\rm as} \quad r\to \infty\, .
\end{cases}
\end{eqnarray}
In Fig.~\ref{fig_d2mfpt1x} we plot $\langle T\rangle_r(1,\, 0)$ vs. $r$ and compare with our
direct numerical simulation, finding excellent agreement. We see once more that as $r\to \infty$,
the MFPT approaches $1$ monotonically given that this starting point $(1,0)$ is a nearest neighbour
of the target at $(0,0)$. Of course, by symmetry, this result in Eq. (\ref{mfpt_d210.2}) 
also holds for three other starting points
$(m_1=0, m_2=1)$, $(m_1=-1, m_2=0)$ and $(m_1=0, m_2=-1)$.

\vskip 0.3cm
\noindent {\bf The case $(m_1=1,m_2=1)$.} In this case, Eq. (\ref{mfpt_d2.1}) reduces to
\begin{equation}
\langle T\rangle_r(1,\, 1)= \frac{1}{r}\left[\frac{2}{\pi}\,
\frac{K\left(\frac{16}{(4+r)^2}\right)}{\int_0^{\infty} dt\, e^{-t}\,
\left[I_{1}\left(\frac{2\, t}{(r+4)}\right)\right]^2}-1\right]\, .
\label{mfpt_d211.1}
\end{equation}
It turns out the integral in the denominator can again be performed using the identity, valid for $|z|<1$,
\begin{equation}
\int_0^{\infty} dt\, e^{-t}\,
\left[I_{1}\left(z\, t/2\right)\right]^2=
\frac{2}{\pi\, z^2}\left[ (2-z^2) K(z^2)- 2\, E(z^2)\right]\, ,
\label{iden.5}
\end{equation}
where $E(u)$ is the EllipticE function with argument $u$ defined as
\begin{equation}
E(u)= \int_0^{\pi/2} \sqrt{1- u\, \sin^2(\theta)}\, d\theta\, .
\label{E_func_def}
\end{equation}
Using the identity \eqref{iden.5} in \eqref{mfpt_d211.1}
leads to the explicit result
\begin{equation}
\langle T\rangle_r(1,\, 1)=\frac{1}{r}\left[\frac{z^2 K(z^2)}{(2-z^2) K(z^2)-2\, E(z^2)}-1\right]\,,
\label{mfpt_d211.2}
\end{equation}
where $z\equiv z(r)= 4/(4+r)$.
The asymptotic behaviors are given by
\begin{eqnarray}
\label{11_asymp}
\langle T\rangle_r(1,\, 1)\approx \begin{cases}
& -\frac{4}{r\, \ln (r)} \quad\quad\quad\quad\quad\quad\,\, {\rm as} \quad r\to 0 \\
\\
& \frac{r}{2}+ 4+ \frac{3}{r}+O(r^{-2}) \quad\, {\rm as} \quad r\to \infty\, .
\end{cases}
\end{eqnarray}
In Fig.~\ref{fig_d2mfpt1x} we plot $\langle T\rangle_r(1,\, 1)$ vs. $r$ and compare with our
direct numerical simulation, finding excellent agreement. We see that as $r\to \infty$,
the MFPT displays a unique minimum at some $r^*$ when this starting point $(1,1)$ is not a nearest neighbour
of the target at $(0,0)$. Again, we note that, by symmetry of the square lattice, the result in Eq. (\ref{mfpt_d211.2}) also
holds for three other points: $(-1,1)$, $(-1,-1)$ and $(1,-1)$.

\subsection{Some explicit results in general dimension $d$}
\label{general_d}

We now consider general dimension $d$ and a general starting point $\vec R_0=a\, (m_1,m_2,\ldots,m_d)$. In this case, 
while the result in Eq. (\ref{mfpt_final}) for the MFPT $\langle T\rangle_r(\vec R_0)$ is exact and easily evaluable 
via numerical integration, it is not easy to reduce it to an explicit formula. However, one can obtain explicitly the 
asymptotics behaviors of the MFPT in the two limits $r\to \infty$ and $r\to 0$, as we show below.

\vskip 0.3cm

\noindent{\bf The limit $r\to \infty$.} We start with the $r\to \infty$ which is easier. In this limit, the argument
$2t/(r+2d)$ of the Bessel functions tend to zero. We can then use the leading asymptotic behavior of $I_{|m|}(z)$ as
$z\to 0$, namely~\cite{Grads},
\begin{equation}
I_{|m|}(z) \approx \frac{1}{\Gamma(|m|+1)}\, \left(\frac{z}{2}\right)^{|m|}\, .
\label{smallz_Bessel.1}
\end{equation}
Substituting this behavior in Eq. (\ref{mfpt_final}) for fixed $(m_1,m_2,\ldots, m_d)$, we get the leading order
behaviors for large $r$
\begin{eqnarray}
\int_0^{\infty} dt\, e^{-t}\,
\left[I_{0}\left(\frac{2\, t}{(r+2d)}\right)\right]^d &\approx & 1  \label{num_smallz.1}
\end{eqnarray}
\begin{multline}
\int_0^{\infty} dt\, e^{-t}\,
    {\displaystyle \prod_{i=1}^d} I_{|m_i|}\left(\frac{2\, t}{(r+2d)}\right)
    \approx  \\
\frac{\Gamma\left({\displaystyle {\sum_{i=1}^d} |m_i| }+1\right)}{ {\displaystyle {\prod_{i=1}^d }} \Gamma(|m_i|+1)}\, 
r^{-(|m_1|+|m_2|+\ldots + |m_d|} \, . \label{den_smallz.1}
\end{multline}
This leads to a rather interesting power law growth of $\langle T\rangle_r(\vec R_0)$
for large $r$ with a distance dependent exponent
\begin{multline}
  \langle T\rangle_r(m_1,m_2,\ldots, m_d)\approx \\
  \frac{{\displaystyle {\prod_{i=1}^d }}\Gamma(|m_i|+1)}{
\Gamma\left({\displaystyle {\sum_{i=1}^d} |m_i| }+1\right)}\, r^{|m_1|+|m_2|+\ldots+|m_d| -1}\, .
\label{large_r_asymp}
\end{multline}
It is easy to check that in $d=1$ and in the special cases for $d=2$, we get back previous results, 
respectively in Eqs. (\ref{mfpt_asymp}), (\ref{10_asymp}) and (\ref{11_asymp}). Note that this result
is completely inaccessible in the continuum limit.

\begin{figure}
\includegraphics[width=0.45\textwidth]{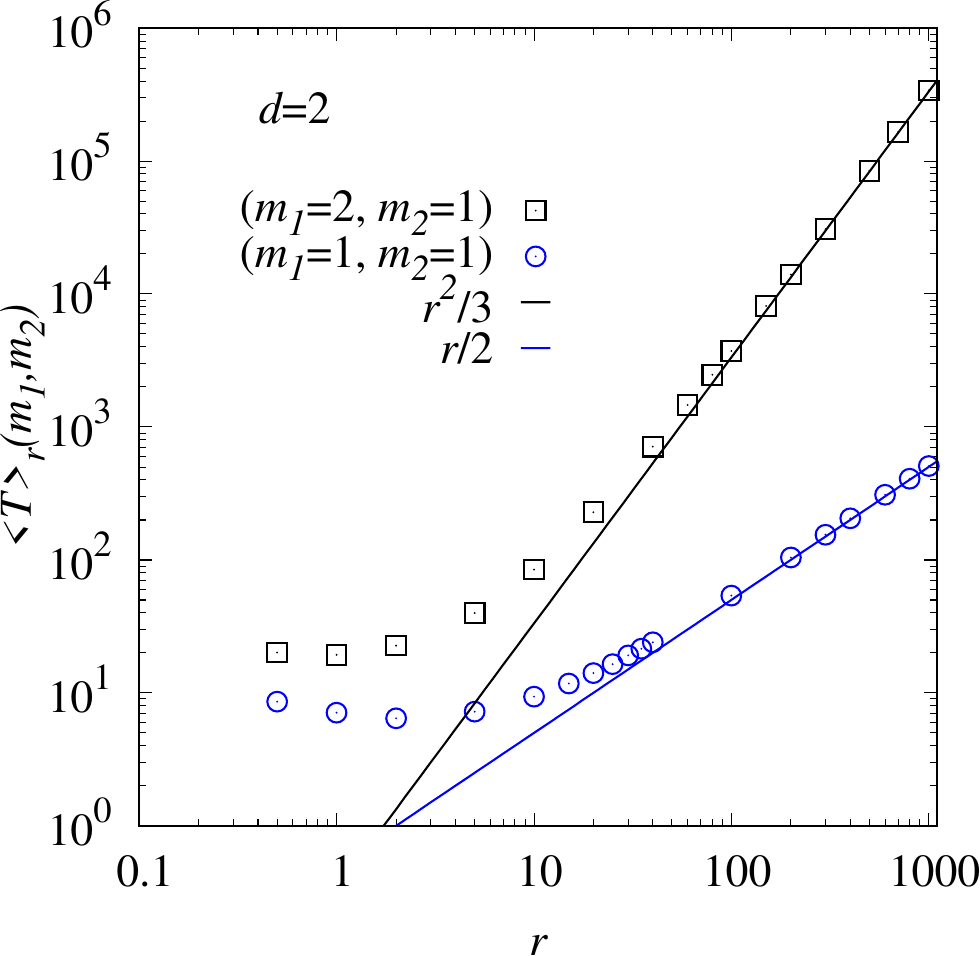}
\caption{$\langle T\rangle_r(1,1)$ vs. $r$ and $\langle T\rangle_r(2,1)$
  vs. $r$ in two dimensions.  The solid lines represent the
    asymptotic large $r$ behavior predicted analytically in
    Eq.\ (\ref{large_r_asymp}), while the symbols represent
the numerical results. The agreement is excellent for large $r$.
}
\label{fig_d2:larger}
\end{figure}

The $r\to \infty$ result in \eqref{large_r_asymp} is valid for any starting point $\vec R_0=a\, (m_1,m_2,\ldots, m_d)$.
Interestingly, this formula predicts that for any starting point which is not a nearest neighbour
of the origin, the MFPT diverges as a power law
$\langle T\rangle_r(m_1,m_2,\ldots, m_d)\sim r^{\phi}$ as $r\to \infty$,
with a distance dependent exponent $\phi=|m_1|+|m_2|+\ldots+|m_d| -1$.
We have checked this result numerically in $d=2$ for $(m_1=1, m_2=1)$
and $(m_1=2,m_2=1)$. In the former case, our result predicts a linear growth with $r$, namely
$\langle T\rangle_r(1,1)\approx r/2$,  while in
the latter case, it predicts a quadratic growth $\langle T\rangle_r(2,1)\approx r^2/3$. 
Numerical results match perfectly our analytical predictions (including the prefactors $1/2$ and $1/3$),
as can be seen in Fig.~\ref{fig_d2:larger}.

In contrast, if the starting point
happens to be one of the $2d$ nearest neighbours of the origin, e.g., for the case $(m_1=1, 0,0,0,\ldots, 0)$
(or any of its $2d$ symmetric cousins), it follows from Eq. (\ref{large_r_asymp}) that 
the MFPT approaches a universal constant $1$ as $r\to \infty$,
\begin{equation}
\langle T\rangle_r(1,0,\ldots, 0)= 1+\frac{2d-1}{r} + O\left(\frac{1}{r^2}\right) \, \quad {\rm as}\quad r\to \infty\, .
\label{rlarge_univ.1}
\end{equation}
We now provide a simple physical argument for this general universal result $1$ as $r\to \infty$.
Consider the walker starting its journey from a nearest neighbour of the origin.
In the limit $r\to \infty$, even if the walker hops
from a nearest neighbour site $\vec R_0$ to any other site (apart from the origin), it immediately
gets reset to $\vec R_0$. Of course, if it jumps to the origin, then the process is over
since the target is found. Thus, in this high resetting limit, the dynamics of the walker
gets essentially localised only to two neighboring sites, namely the starting site $\vec R_0$
(nearest neighbour to the origin) and the origin itself. The effective dynamics in continuous time
then reduces to the following: in a small time interval $dt$, the walker hops from
$\vec R_0$ to $\vec 0$ with probability $dt$ and with the complementary probability
$(1-dt)$, it stays at the departure site $\vec R_0$. Thus it becomes a Poisson process
with rate $1$. The probability that the walker stays at the starting site $\vec R_0$ up to time $t$
is simply $Q(\vec R_0,t)=e^{-t}$. Consequently, the first-passage probability density to the origin, starting from
$\vec R_0$, is simply $F(\vec R_0, t)= -\partial_t Q(\vec R_0,t)= e^{-t}$. Hence the MFPT is given by its first moment
\begin{equation}
\lim_{r\to \infty} \langle T\rangle_r (\vec R_0)= \int_0^{\infty} t\, e^{-t}\, dt=1\, .
\label{mfpt_univ.1}
\end{equation}
This argument is very general and leads to this universal limiting value $1$ in any dimension,
as long as the starting point $\vec R_0$ is a nearest neighbour of the target. 
\change{A similar effective two-site model can be constructed even for discrete-time random walker when
the starting site is a nearest neighbour of the origin.}
The fact that
the global minimum of the MFPT occurs at $r=r^*=\infty$ when the starting point is a nearest neighbour
of the target is another striking result that is not captured in the continuum limit, also 
simply because there exists no nearest neighbour in the continuum
limit.

\vskip 0.3cm

\noindent{\bf The limit $r\to 0$.} This limit, for general $\vec R_0=a\,
(m_1,m_2,\ldots, m_d)$,  
turns out to be more tricky and the behaviors depend on the dimension $d$.

\begin{itemize}

\item $d=1$. In this case, from the explicit result in Eq. (\ref{mfpt_asymp}), we have the leading order 
result as $r\to 0$
\begin{equation}
\langle T\rangle_r (m_1) \approx \frac{|m_1|}{\sqrt{r}}\, ,
\label{d1r0.1}
\end{equation}
valid for arbitrary $m_1$.

\item $d=2$. In this case, we had exact results only in two cases:
for the starting points $(m_1=1, m_2=0)$ (and its $3$ symmetric counterparts) and $(m_1=1, m_2=1)$ (along with
its three symmetric cousins).
In both cases, the MFPT diverges as $\langle T\rangle_r\approx -A/(r\, \ln r)$ as $r\to 0$ where the prefactor
$A=\pi$ in the first case (see Eq. (\ref{10_asymp})), while $A=4$ in the second case (see Eq. (\ref{11_asymp})).
We now show that for general $(m_1, m_2)$, the MFPT diverges as $r\to 0$ exactly in the same way, but with a
prefactor $A(m_1,m_2)$ that depends explicitly on the starting point.
More precisely, as $r\to 0$, we get
\begin{equation}
  \langle T\rangle_r(m_1,m_2) \approx - \frac{A(m_1,m_2)}{r\, \ln r}\, ,
\end{equation}
where
\begin{multline}
  A(m_1,m_2)= \\\int_0^{\infty} dt\,
  e^{-t}\, \left[ I_0^2(t/2)- I_{|m_1|}(t/2)\, I_{|m_2|}(t/2)\right]\, .
  \label{d2r0.1}
\end{multline}
To prove this result, we start from our exact result in Eq. (\ref{mfpt_d2.1}) valid for arbitrary $(m_1,m_2)$ that
reads
\begin{widetext}
\begin{equation}
\langle T\rangle_r(m_1,\, m_2)=\frac{1}{r}\, \left[ \frac{ \int_0^{\infty} dt\, e^{-t}\,
\left[ I_{0}^2\left(\frac{2\, t}{(r+4)}\right)^2 - I_{|m_1|}\left(\frac{2\, t}{(r+4)}\right)\, 
I_{|m_2|}\left(\frac{2\, t}{(r+4)}\right)\right]}
{\int_0^{\infty} dt\, e^{-t}\,
I_{|m_1|}\left(\frac{2\, t}{(r+4)}\right)\, I_{|m_2|}\left(\frac{2\, t}{(r+4)}\right)}\right]\, .   
\label{mfpt_d2.2}
\end{equation}
\end{widetext}
Let us first consider the integral in the denominator  
\begin{equation}
F_r(m_1,m_2)= \int_0^{\infty} dt\, e^{-t}\, I_{|m_1|}\left(\frac{2}{r+4}\,t\right)\, I_{|m_2|}\left(\frac{2}{r+4}\,t
\right)\, .
\label{Fz.1}
\end{equation}
We want to extract its behavior as $r\to 0$. By making a
change of variable $2t/(r+4)=y$, we get
\begin{multline}
  F_r(m_1,m_2)=\\
  \frac{(r+4)}{2}\, \int_0^\infty dy\, e^{-ry/2}\, e^{-2y}\, I_{|m_1|}(y)\, I_{|m_2|}(y)\, .
\label{Fz.2}
\end{multline}
We note that since $I_{|m|}(y)\approx e^{y}/\sqrt{2\pi y}$ as $y\to \infty$ for any $|m|$, it is clear that the
integrand in \eqref{Fz.2} behaves as $e^{-ry/2}/(2\pi y)$ for large $y$ and hence the integral diverges as $r\to 0$.
To extract this leading divergence, we divide the integral over $y$ into two parts: $y\in [0,\Lambda]$
and $[\Lambda, \infty]$ where the cutoff $\Lambda\gg 1$ but is independent of $r$ as $r\to 0$. Then the leading divergence
as $r\to 0$ comes from the second interval $y\in [\Lambda,\infty]$ where we can use the asymptotic behavior of 
the Bessel function $I_{|m|}(y)\approx e^{y}/\sqrt{2\pi y}$. This gives
\begin{eqnarray}
  F_r(m_1, m_2) & \approx &  2\, \int_{\Lambda}^{\infty} dy\,  \frac{ e^{-ry/2}}{2\pi y} \\
  &= &  
  \frac{1}{\pi}\, 
\int_{\Lambda\, r/2}^{\infty} 
dz\, \frac{e^{-z}}{z}\approx
-\frac{1}{\pi}\, \ln r 
\label{Fz.3}
\end{eqnarray}
as  $r\to 0$.
We can now use this behavior for the denominator in Eq. (\ref{mfpt_d2.2}) and set $r=0$ in the numerator (since
the integral in the numerator is convergent as $r\to 0$). This then gives the leading order $r\to 0$ behavior in
Eq. (\ref{d2r0.1}).

\item $d>2$. We start from our general result \eqref{mfpt_final}
  which is valid for any $r$ and any starting point
  $\vec R_0=a\vec m = a \, (m_1,m_2,\ldots, m_d)$.
In this case,  by using the asymptotic behavior $I_{|m|}(y)\approx e^{y}/\sqrt{2\pi y}$, 
it is easy to check that both the integrals in the numerator and the denominator
are convergent as $r\to 0$. Hence the leading $r\to 0$ behavior of the MFPT is clearly
then given by 
\begin{equation}
  \langle T\rangle_r(\vec m)\approx\\
  \frac{B(\vec m)}{r}\,, 
  \label{highd.1}
\end{equation}
as $ r\to 0$ and where the prefactor is
\begin{eqnarray}
  B(\vec m) & \equiv & B(\vec R_0) \nonumber \\
  & = &
\frac{\int_0^{\infty} dt\, e^{-t}\, I_0^{d}\left(\frac{t}{d}\right)}{ 
\int_0^{\infty} dt\, e^{-t}\, {\displaystyle {\prod_{i=1}^d }} I_{|m_i|}\left(\frac{t}{d}\right) }-1\, .
\label{highd_amp.1}
\end{eqnarray}
Actually this prefactor $B(\vec R_0)$ in Eq. (\ref{highd_amp.1}) has a nice physical interpretation.
To extract this physical meansing, let us go back to the general expression in Eq. (\ref{mfpt_fp.1}).
We see in Eq. (\ref{mfpt_fp.1}) that in the limit $r\to 0$ and for $d>2$, one has
\begin{eqnarray}
  \langle T\rangle_r(\vec R_0) & \approx&
  \frac{1}{r}\, \left [\frac{1}{\tilde{F}_0(\vec R_0,0)}-1\right] \nonumber\\
  & = &
\frac{1}{r}\, \left[ \frac{1}{\int_0^{\infty} F_0(\vec R_0, t)\, dt}-1\right]\, .
\label{highd_phys.1}
\end{eqnarray}
Comparing with Eq. (\ref{highd.1}), we then identify the amplitude $B(\vec R_0)$ as
\begin{equation}
B(\vec R_0)= \left[ \frac{ 1-\int_0^{\infty} F_0(\vec R_0, t)\, dt}{\int_0^{\infty} F_0(\vec R_0, t)\, dt}\right]\, .
\label{B_amp.1}
\end{equation}
Now the denominator $\int_0^{\infty} F_0(\vec R_0, t)\, dt <1$ is just the hitting probability of
the target in $d>2$.
We recall that for an ordinary random walker without resetting in $d>2$, the target gets captured by
the walker
with probability $\int_0^{\infty} F_0(\vec R_0, t)\, dt <1$, while with the complementary
probability $1- \int_0^{\infty} F_0(\vec R_0, t)\, dt $ the walker escapes to infinity.
Hence, from Eq. (\ref{B_amp.1}), we see that the prefactor $B$ is just the ratio of the
escape and the hitting probability for a walker in $d>2$ starting at $\vec R_0$
\begin{equation}
B(\vec R_0)= \frac{ {\rm escape}\, {\rm probability}}{ {\rm hitting}\, {\rm probability}}\, .
\label{B_phys.1}
\end{equation} 
Here, we thus provide an explicit expression for this ratio in Eq. (\ref{highd_amp.1}) in terms
of integrals over Bessel functions.

\end{itemize}

\vskip 0.3cm

\noindent {\bf Optimal resetting rate in the large dimension limit $d\to \infty$.} From the discussion above,
it is clear that as long as the starting point is not a nearest neighbour of the target at the origin, the MFPT
$\langle T\rangle_r(\vec R_0)$, as a function of $r$ for fixed $\vec R_0$, diverges at the two limits
$r\to 0$ and $r\to \infty$, displaying a single minimum at the optimal value $r^*$ which depends on both
the starting position $\vec R_0$ and dimension $d$. For a fixed starting position $\vec R_0$, it is
interesting to ask how $r^*(d)$ behaves as a function of dimension $d$? To determine $r^*$ explicitly
from the exact expression \eqref{mfpt_final} seems difficult. However, one can make progress in
the large dimension limit. In this limit, the factor $2/(r+2d)$ in the argument of the Bessel functions
in \eqref{mfpt_final} becomes small and one can make a systematic large $d$ expansion of \eqref{mfpt_final}
by using the asymptotic behavior of Bessel function for small arguments given in Eq.\ (\ref{smallz_Bessel.1}).
This systematic large expansion of Eq.\ (\ref{mfpt_final}) yields, for fixed $r$ and $\vec R_0=a (m_1,m_2,\ldots, m_d)$,
the following expression for the MFPT up to leading orders in $d$
\begin{multline}
  \langle T\rangle_r(m_1,m_2,\ldots, m_d)\approx \\
  \beta\, (2d)^{\alpha}
\Bigg[    \frac{1}{r} + \frac{\alpha\, r+1}{r}\, \frac{1}{2d}
    + \\
    \left(\alpha-2 + \frac{\alpha(\alpha-1)}{2}\, r\right)\,
    \frac{1}{(2d)^2} + O\left(\frac{1}{d^3}\right) \Bigg]\, ,
\label{larged.1}
\end{multline}
where
\begin{equation}
\alpha= {\displaystyle {\sum_{i=1}^d}} |m_i| \, \quad {\rm and}\quad
\beta = \frac{ {\displaystyle {\prod_{i=1}^d}} \Gamma(|m_i|+1)}{\Gamma(|m_1|+|m_2|+\ldots+|m_d|+1)}\, .
\label{ab_def}
\end{equation}
One can now minimize \eqref{larged.1} with respect to $r$ and one gets
\begin{equation}
r^*(d, (m_1,m_2,\ldots, m_d))\approx  \sqrt{\frac{\change{8}}{\alpha(\alpha-1)}}\, d\, , \quad {\rm as}\quad d\to \infty\, .
\label{r_opt_d}
\end{equation}
Let us recall that since the starting point is not a nearest neighbour of the origin, we must have 
$\alpha>1$. In particular, if the starting point happens to be
$(2,0,0,\ldots, 0)$, then $\alpha=2$ and we have from Eq.~(\ref{r_opt_d}) 
\begin{equation}
r^*(d, (2, 0,0,\ldots, 0)) \approx 2\, d\, ,
\label{r_opt_2}
\end{equation}

\begin{figure}
\includegraphics[width=0.45\textwidth]{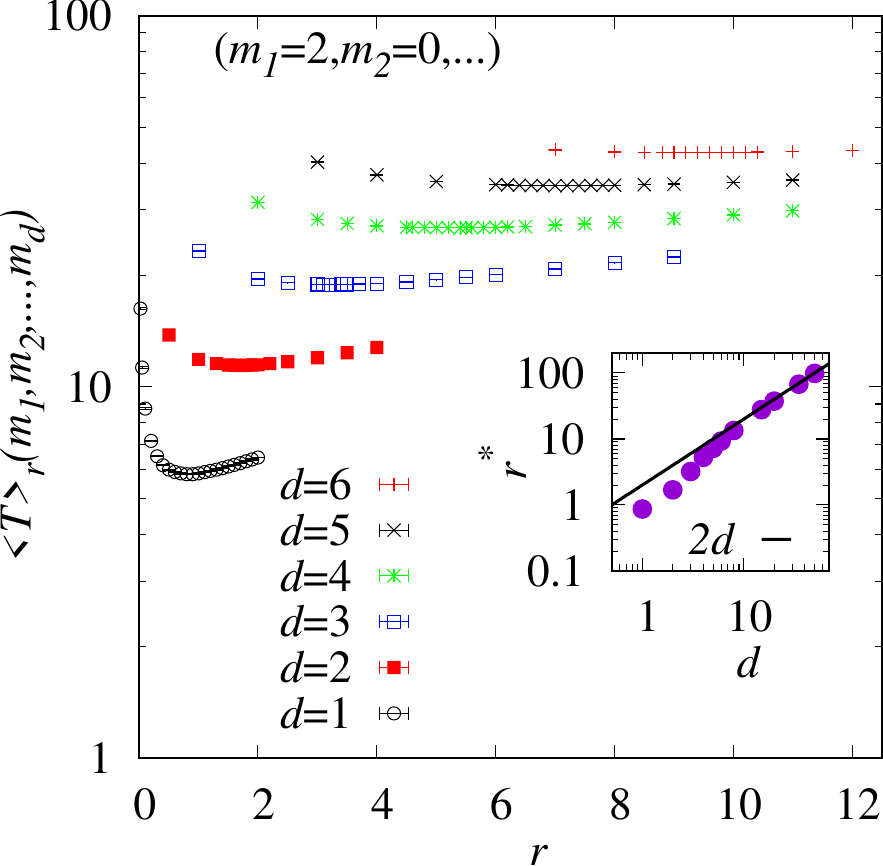}
\caption{The simulation results
  $\langle T\rangle_r(2,0,0,\ldots, 0)$ vs. $r$ in dimensions $d=1$
  to $d=6$. The minima represent the optimal resetting rate $r^*$.
  The inset shows $r^*$ as a function of $d$ for some values $d\le 50$
  which seems to fit at larger value of $d$ well with
  the linear growth $r^*(d)=2 d$ predicted in Eq.\ (\ref{r_opt_2}). 
} 
\label{fig:MFPT_d}
\end{figure}

This result agrees well with our numerical simulations, as shown in
Fig.~\ref{fig:MFPT_d}. Here we have determined
$\langle T\rangle_r(2,0,0,\ldots,0)$ for various
dimensions $d\in[1,50]$ and selected values of $r$. We have determined
the optimum resetting rates $r^*$ and minimum first-passage times $T_0$
by fitting parabolas
$\tilde T(r)=a(r-r^*)^2+T_0$ near the minimum of
$\langle T\rangle_r(2,0,0,\ldots,0)$, respectively.
The inset of the plot shows $r^*$ as function of dimension $d$, the linear
growth for large $d$ is well visible.

\begin{figure}
\includegraphics[width=0.45\textwidth]{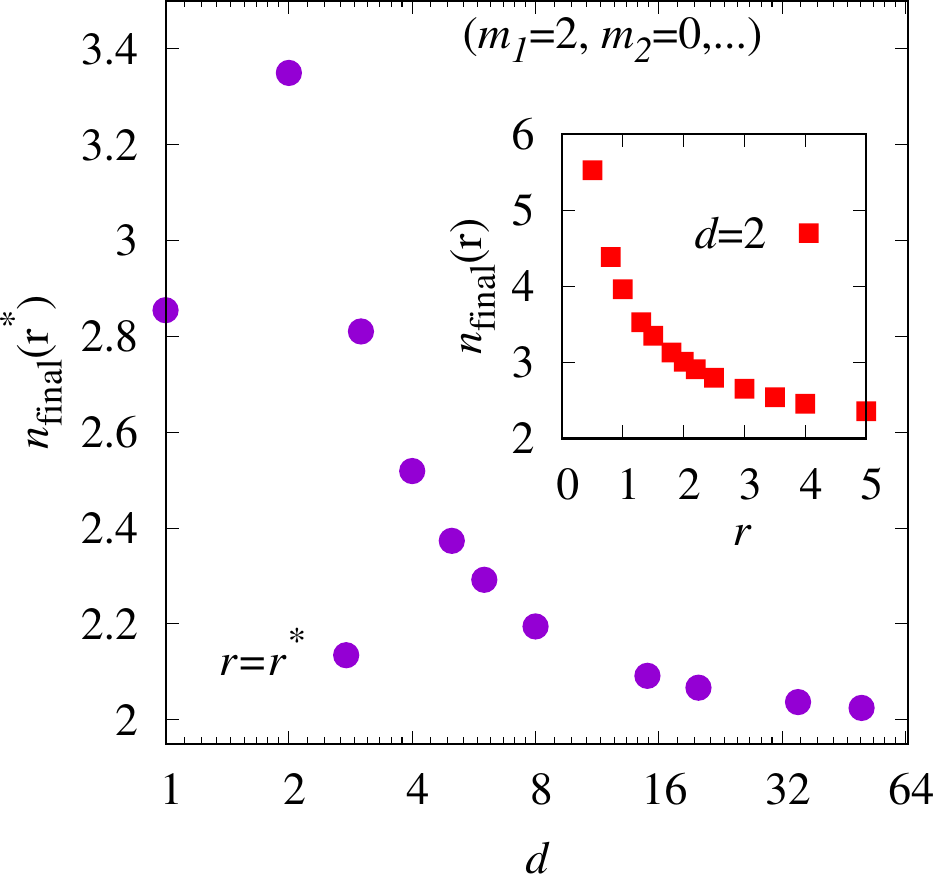}
\caption{The number $n_{\rm final}(r^*)$ of steps the walker takes since the final,
  i.e.\ successful, reset until reaching the target, measured at the
  optimium reset rate $r^*$, shown as function of the lattice dimension $d$.
  Here the target is located at $m=(2,0,\ldots,0)$.
  The inset shows $n_{\rm final}$ as function of resetting rate $r$ for dimension
$d=2$.
}
\label{fig:n_walker}
\end{figure}

In our simulations, we have also investigated the paths
to the target after the last reset. 
With increasing dimension, at each position,
the number of possible steps leading away from
the target increases. Thus, the walker needs to be reset
more and more to find the target as the dimension increases.
This also indicates that the actual paths leading to the target after
the last reset will more and more be shortest paths. We have verified
this for the starting position $\vec m=(2,0,\ldots,0)$ by measuring the
number of steps $n_{\rm final}$ the walker takes to reach the target
after the last reset has happened.
In Fig.~\ref{fig:n_walker}
we show the value $n_{\rm final}(r^*)$ for the optimum resetting rate $r^*$
as function of dimension $d$. Indeed
one observes an convergence of $n_{\rm final}$ to the value 2. Note that
also for increasing $r$ this convergence to the value $n_{\rm final}=2$
is visible, see inset of Fig.~\ref{fig:n_walker}. This also makes sense,
because with increasing resetting rate, the walker has less time
to explore the space, i.e., only the shortest paths to the target
will occur.

Note that we have also measured the total number $n_{\rm hops}$
of hops since $t=0$, see Section \ref{sec:algorithm},
but this correlates almost perfectly with
$\langle T\rangle_r(2,0,\ldots,0)$, so we do not
show these results here.

The observation that with increasing lattice dimension $d$, or
increasing reset rate $r$, only direct
paths to the target dominate, motivates a simple finite-state
Markov chain model, which allows
one to derive the optimum resetting rate for close targets very quickly.
Since this is pedagogically interesting, we also present this approach here.

We quickly recall from the theory of Markov chains \cite{grinstead2012} how to
calculate the mean-first passage time.
For simplicity, we consider a discrete time
Markov chain with $N+1$ states, where the state $N+1$ is the
single absorbing state. Let the column vector
$\vec{\pi}(s)=(\pi_1(s),\ldots,\pi_{N+1}(s))^T$ denote
the probabilities that
the chain is in state $j$ at step $s$. The dynamics of the Markov
chain is described by transition matrix $\mat{P}$ according to
$\vec{\pi}(s+1)=\mat{P} \vec{\pi}(s)$, where $p_{ij}$ is the probability that
the chain moves to state $i$ if it is in state $j$.
Since state $N+1$ is absorbing,
we can write $\mat{P}$ as
\begin{equation}
  \mat{P}=\left(\begin{array}{cc}
    \mat{Q} & \vec{0} \\
  \vec{u}^T  & 1
  \end{array}\right)\,,
\end{equation}
where the $N\times N$ matrix $\mat{Q}$ describes the transition
probabilities excluding the absorbing state, i.e., among the transient states.
The vector $\vec{u}$ denotes
the transition probabilities to the absorbing state. Now, the
matrix $\mat{P}^n$ describes the dynamics of $n$ steps, i.e. its element
$p^{(n)}_{ij}$ the probability that the chain is after $n$ steps in state $i$
if it starts in state $j$. Correspondingly, for the matrix $\mat{Q}^n$, does
the same for the set of transient states. This leads to the definition
of the fundamental matrix
\begin{equation}
\mat{N} = \mat{I}+\mat{Q}+\mat{Q}^2+\cdots \,,
\end{equation}
where $\mat{I}$ is the identity matrix. 
The entry $n_{ij}$ describes the expected number of times
the chain has visited transient state $i$,
given that the chains has started in state $j$.
Therefore,
the mean-first passage steps, starting at state $j$ is given the expected
number of times it has visited any transient state, i.e., by
\begin{equation}
  \tilde T=\sum_{i=1}^N n_{ij}\,.
  \label{eq:fpt:markov}
\end{equation}
$\mat{N}$ can be
conveniently calculated via the matrix inverse
$\mat{N}=(\mat{I}-\mat{Q})^{-1}$,
reminiscent of the well-known geometric-series
identity $\sum_{k=0}^\infty q^k = \frac 1 {1-q}$ for $|q|<1$.

\begin{figure}
  \includegraphics[width=0.18\textwidth]{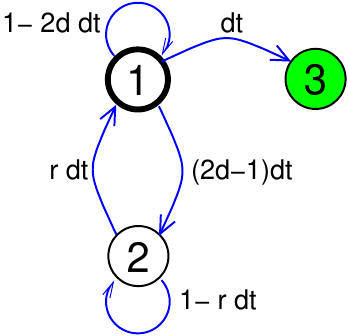}
\includegraphics[width=0.27\textwidth]{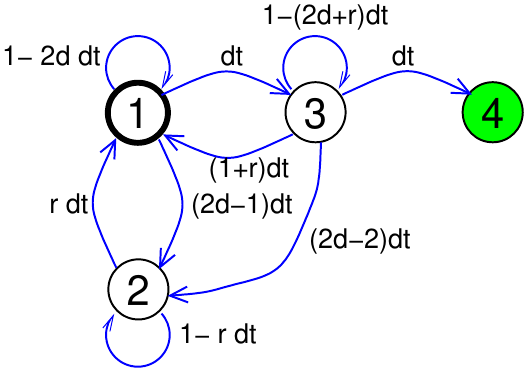}
\caption{Markov chains for representing lattices in $d$ dimensions
  where (left) the target site (state 3)
  is located next to the starting and resetting site (state 1),
  or (right) where they are separated in one lattice direction by an
  intermediate site (state 3). State 2 represents all other sites.
  It is assumed, which holds for $d\to \infty$ or $r\to \infty$,
  that all relevant paths to the target site do not go through state 2.
}
\label{fig:markov_chains}
\end{figure}

We start with the case where the target is next to the starting
position $\vec R_0$,
which is also the resetting position.
Here we approximate the dynamics by three states: state 1 is the starting and
resetting site, state 3 the target site, and state 2 represents all
other sites. Since we assume that only direct paths from the starting
site to the target sites are relevant, we do not consider transitions
between states 2 and 3. To work in probability space, we assume a small
time step $dt$, thus, resetting occurs with probability $rdt$,
and a move to each  of the $2d$ neighbouring sites with rate $1$, i.e.,
probability $dt$. Most sites of state 2 are not neighbors of the
site $\vec R_0$, so we also ignore transitions to state 1 except resetting.
The states and all transitions are displayed in Fig.\ \ref{fig:markov_chains}.
The corresponding transition matrix reads as follows:

\begin{equation}
  {\bf P} = \left(
  \begin{array}{rrr}
    1-2d\,dt & r dt & 0 \\
    (2d-1) dt & 1- rdt & 0 \\
    dt & 0 & 1
  \end{array}
  \right)\,,
\end{equation}
where ${\bf Q}$ is the upper left $2\times 2$ matrix. The calculation of
${\bf N}$ can be performed simply  by using an symbolic linear algebra package.
The calculation of the first-passage time via
Eq.~(\ref{eq:fpt:markov}) and converting to continuous
time by $T=\tilde T dt$
yields
\begin{equation}
  \langle T \rangle_r(1,0,\ldots)=  1+\frac{2d-1}{r} \quad (d\to\infty, r \to \infty)
\end{equation}
Thus, the optimum resetting rate is $r\to\infty$, which means the
walker stays on the starting site until it moves in one step to the
target site.
This convergence to 1 matches the $r\to\infty$ behavior
from Eq.~(\ref{rlarge_univ.1}), which is
$d=2$  functional form
displayed in Eq.~(\ref{10_asymp}).

When having the starting site $\vec m=(2,0,\ldots,0)$,
represented again by state 1,
two sites away from the target site, here state 4,
we have to include $\vec m'=(1,0,\ldots,0)$ as
intermediate site explicitly, here state 3.
The walker will move with rate 1 from the
starting site to $\vec m'$ with rate $1$. It can move back either
by a hop with rate $1$ or reset with rate $r$, yielding $1+r$
as total rate. It may move to the target site with rate 1. Thus, it may move to
other neighbouring sites, represented by state 2,
with rate $2d-2$. Again, not considering transitions between all other
sites and the target or intermediate sites, and transforming the rates
to probabilities by multiplying with $dt$,
the transition matrix reads as follows:

\begin{equation}
  {\bf P} = \left(
  \begin{array}{rrrr}
    1-2d\, dt & rdt& (1+r) dt & 0 \\
     (2d-1) dt &    1-rdt  & (2d-2) dt& 0\\
     dt &  0 & 1- (r+2d) dt &0 \\
     0 &  0 dt && 1
  \end{array}
  \right)\,,
\end{equation}
Obtaining ${\bf Q}$, inverting ${\bf I-Q}$ and getting
$\tilde T dt$ via Eq.~(\ref{eq:fpt:markov}) results in
\begin{equation}
  \langle T \rangle_r(2,0,\ldots)=  r+4d+\frac {4d^2-2}{r}
  \quad (d\to\infty, r \to \infty)
\end{equation}
For $d=2$ this yields $r+8+6/r$ which is twice the result of
Eq.~(\ref{11_asymp}) for $\vec m=(1,1)$, which makes sense, because
for that case there are two shortest path to the target, i.e., half the
time is needed on average.
This function exhibits a minimum, which is obtained by setting
$\frac{d T}{dr}(r^*)=0$,
i.e., $r^* =\sqrt{4d^2-2} \to 2d$ for large values of the dimension $d$,
compatible with the previous result of Eq.~(\ref{r_opt_2}).

In summary, we find that in general dimension $d$, if the starting point $m$
is not a nearest neighbour of 
the target at the origin, then the MFPT in Eq. (\ref{mfpt_final}), as a function of $r$, diverges at both ends 
$r\to 0$ and $r\to \infty$, and displays a minimum at an optimal value $r^*(\vec R_0)$ that depends expliicitly on 
the starting point $\vec R_0$ and the dimension $d$. In Fig.~\ref{fig:MFPT_d}
we show the MFPT as a function of $r$, 
for the starting point $\vec R_0= (m_1=2, m_2=0, m_3=0,\ldots, m_d=0)$, for dimensions $d=1,2,3,4, 5$. We see that $r^*$ 
increases with increasing $d$. In principle, one can numerically determine this optimal value by setting to zero 
the derivative of Eq. (\ref{mfpt_final}) with respect to $r$ and $r=r^*$ and then determining this root $r^*$ 
numerically. For example, for $d=1$ and with $m_1=2$, this gives $r^*\approx 0.828489$ which 
matches very well with the simulation value. We have shown that in the large $d$ limit, one
can prove analytically that $r^*(d)\approx 2 d$ when the starting position is  
$\vec R_0= (m_1=2, m_2=0, m_3=0,\ldots, m_d=0)$.
In contrast, if the 
starting point happens to be a nearest neighbour of the target, then the MFPT decreases monotonically with 
increasing $r$ and achieves its lowest value $1$ as $r\to \infty$,
see Eq.\ (\ref{rlarge_univ.1}). Hence, in this 
case, the optimal value $r^*$ is infinite, a result somewhat unexpected.

\section{Nonequilibrium stationary state in the absence of a target}
\label{NESS_section}

\begin{figure}
\includegraphics[width=0.45\textwidth]{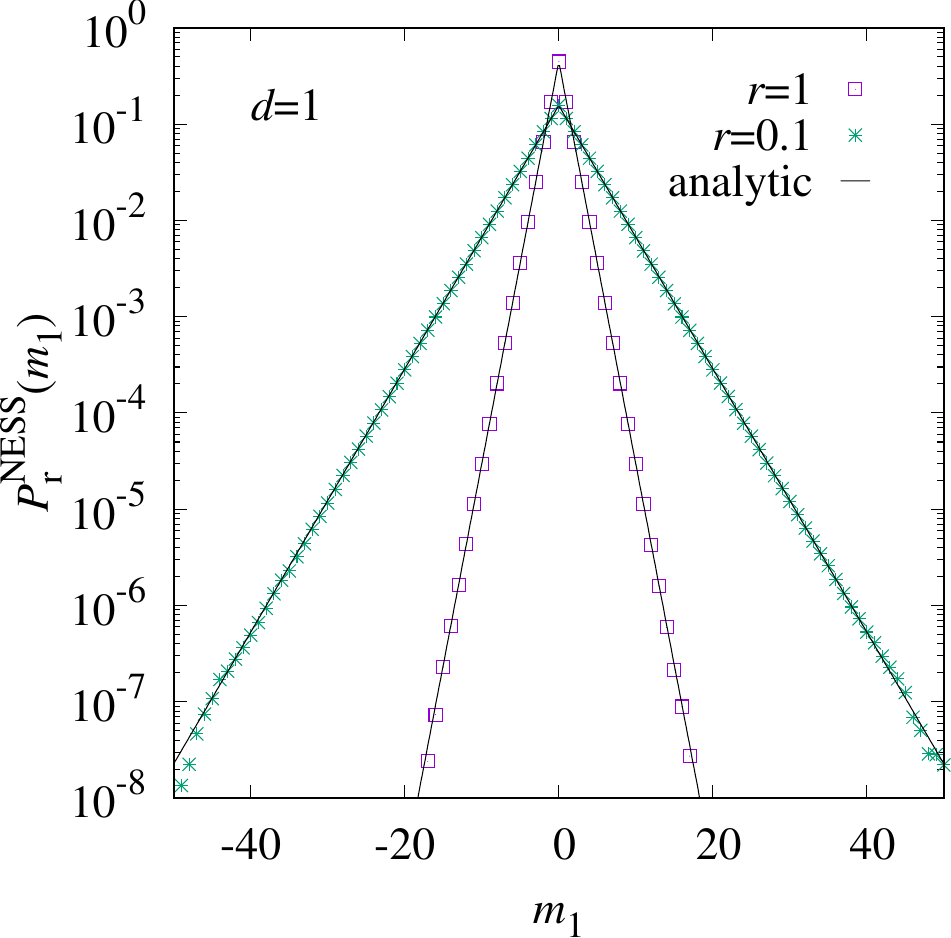}
\caption{Non-equilibrium steady-state density for dimension $d=1$
  from Eq.\ (\ref{stat_pd_d1}), shown as lines, compared to the
  numerically obtained density, shown as symbols, for two cases $r=1$ and
$r=0.1$. 
}
\label{fig:density_1d}
\end{figure}

So far, we have focused on the MFPT to a target placed at the origin for a resetting random walker on a $d$-dimensional lattice.
In the absence of a target, as in the continuous space case, the resetting to the initial positio $\vec R_0$ with
a constant rate $r$ drives the system into a nonequilibrium stationary state (NESS) at long times. In this section,
we present the position distribution in the NESS for such a walker on the lattice. Since there is no target, the system
is translationally invariant on the lattice, and without any loss of generality, we can set the starting (and the resetting)
position $\vec R_0=\vec 0$. Then we just need to compute $P_r(\vec R, \vec 0, t)$ in the long time limit.
This probability evolves via the Fokker-Planck equation \eqref{fp.1} discussed in Section \ref{Model}.
One can easily solve this linear equation. However, the solution can also be obtained by employing a
renewal equation approach. One can express the probability in the preence of resetting in terms
of the probability without resetting via
\begin{equation}
P_r(\vec R, \vec 0, t) = e^{-r\, t}\, P_0(\vec R, \vec 0, t) + r\, \int_0^{\infty} d\tau_l\, e^{-r\, \tau_l}\, 
P_0(\vec R, \vec 0, \tau_l)\, .
\label{pd_renewal.1}
\end{equation}
The first term describes the `no resetting' event in the interval $[0,t]$ with the walker freely propagating
from $\vec 0$ to $\vec R$ in time $t$ without resetting. The second term describes events with one or more resettings.
In this case, let $t-\tau_l$ denotes the epoch at which the last resetting event took place before $t$. Then
$\tau_l$ is simply the time interval between $t$ and the last resetting event before $t$. Since after the last resetting,
the walker propagates freely from time $t-\tau_l$ to $t$, we have the factor $P_0(\vec R, \vec 0, \tau_l)$ inside
the integral on the rhs of \eqref{pd_renewal.1}. The probability that there is no resetting event in the interval
$[t-\tau_l]$ and $t$, preceeded by a resetting event just at $t-\tau_l]$ is simply $r\, e^{-r\, \tau_l}\, d\tau_l$.
Multiplying the two and integrating over all possible values of $\tau_l\in [0,t]$ gives the second term in
\eqref{pd_renewal.1}. One can check that the solution \eqref{pd_renewal.1} indeed satisfies the Fokker-Planck
equation (\ref{fp.1}) at all times with the correct initial and boundary conditions. Finally, in the limit $t\to \infty$
the first term drops out and we obtain the stationary position distribution in the NESS
\begin{multline}
  P_r^{\rm NESS}(\vec R)= P_r(\vec R, \vec 0, \infty)= \\
  r\, \int_0^{\infty} d\tau_l\, e^{-r\, \tau_l}\,
P_0(\vec R, \vec 0, \tau_l)= r\, \tilde{P}_0(\vec R, \vec 0, r)\, ,
\label{stat_pd.1}
\end{multline}
 where $\tilde{P}_0(\vec R, \vec 0, r)$ is just the Laplace transform of the position distribution without resetting
and has already been computed explicitly in \eqref{int_rep.3}. Using this result in \eqref{stat_pd.1} gives
us the exact position distribution in the NESS
\begin{multline}
  P_r^{\rm NESS}(\vec R= a\, (m_1,m_2,\ldots, m_d))=\\
  \frac{r}{r+2d}\, \int_0^{\infty} dt\, e^{-t}\,
{\displaystyle \prod_{i=1}^d} I_{|m_i|}\left(\frac{2\, t}{(r+2d)}\right)\, .
\label{stat_pd.2}
\end{multline}
For example, in $d=1$, using the identity \eqref{iden.2}, we get
\begin{equation}
P_r^{\rm NESS}(R= a\, m_1)= \frac{r}{\sqrt{r^2+4\, r}}\, \left[\frac{r+2-\sqrt{r^2+4\, r}}{2}\right]^{|m_1|}\, ,
\label{stat_pd_d1}
\end{equation}
recovering the $d=1$ result obtained in Ref.~\cite{C2021}.
One can check, by summing over $m_1$, that $P_r^{\rm NESS}(R)$ is
normalized to unity. Eq.\ (\ref{stat_pd_d1}) is compared in
Fig.~\ref{fig:density_1d} to the numerical density for resetting rates
$r=0.1$ and $1$, respectively. The numerical results
were obtained from $10^6$ independent runs up to time $t=10^3$,
respectively.

The expression in Eq. (\ref{stat_pd.2}) is the main result of this section. Following exactly the same method as
in Section \ref{lattice} and the Appendix \ref{Cont_lim_App}, one can show that in the limit of lattice spacing $a\to 0$,
one recovers the continuum results found in Ref.~\cite{EM2014} in general dimension $d$. We do not repeat the calculation here.
For example, in $d=1$, one obtains by taking $a\to 0$ limit in Eq. (\ref{stat_pd_d1})
\begin{equation}
P_r^{\rm NESS}(x= a\, m_1) \to \frac{a}{2}\, \sqrt{\tilde{r}}\, e^{- \sqrt{\tilde{r}}\, |x|}\, ,
\label{stat_d1}
\end{equation}
where $\tilde{r}= r/a^2\, $ is the rescaled resetting rate. Using the fact that the lattice position distribution
is just $a$ times the position density in the continuum, one then recovers the well known result in the continuum
in $d=1$ (upon setting the diffusion constant $D=1$)~\cite{EM2011}. 
Note however that, as in the case of the MFPT in earlier sections, the continuus space result requires
taking the scaling limit $r\to 0$, $R\to \infty$, while keeping the product $\sqrt{r}\, R$ fixed.
This is true in all dimensions. However, the lattice result in \eqref{stat_pd.2} is valid for
arbitrary $r$ and arbitrary $\vec R$ and hence has richer informations that are not accessible
within the continuum approach.

As an example, we consider again the limit $r\to \infty$, i.e., large resetting rate. Taking this limit
in Eq. (\ref{stat_pd.2}) by following the same procedure as in Section \ref{general_d}, one gets
\begin{multline}
P_r^{\rm NESS}(\vec R= a\, (m_1,m_2,\ldots, m_d))\approx \\
\frac{\Gamma\left({\displaystyle {\sum_{i=1}^d} |m_i| }+1\right)}{ {\displaystyle {\prod_{i=1}^d }} \Gamma(|m_i|+1)}\,
r^{-(|m_1|+|m_2|+\ldots + |m_d|)}\, .
\label{larger_stat.1}
\end{multline}
This means that as one goes further away from the origin, the position distribution decays exponentially
as $\sim \exp\left[- \ln r\, (|m_1|+|m_2|+\ldots+|m_d|)\right]$ and the decay length scales for large $r$
as $\sim 1/(\ln r)$. This large $r$ result can not be captured in the continuum limit.

\section{Conclusion}
\label{Conclusion_section}

In this paper we have studied a single particle diffusing on a $d$-dimensional lattice starting from a fixed 
initial position $\vec R_0$ and stochastically resetting to $\vec R_0$ with rate $r$. Our main focus was on the 
mean first-passage time (MFPT) to a target at the origin, as a function of the resetting rate $r$ and the initial 
position $\vec R_0$. We first derived a general formula relating the MFPT to the lattice Green's function and then 
using it, found an exact formula for the MFPT on a $d$-dimensional hypercubic lattice. The previous known results 
were for diffusion in continuous space which we recover from our more general formula in the scaling limit: 
$r\to 0$, $R_0\to \infty$, but keeping $\sqrt{r}\, R_0$ fixed. However, our exact formula
lets us explore a much wider parameter space, i.e., for different values of $r$ as well as $\vec R_0$, going
beyond the continuum theory.
One such interesting result is that the MFPT, as a function of $r$ for fixed $\vec R_0$, diverges as 
$r\to 0$ and $r\to \infty$ with a minimum in between, provided the starting point is not a nearest neighbour of 
the target. In this case, the MFPT diverges as a power law as $r\to \infty$, i.e.,
$\langle T\rangle_r (\vec R_0)\sim r^{\phi}$,
but with an exponent $\phi= (|m_1|+|m_2|+\ldots+ |m_d|)-1$ that varies with 
the starting position $\vec R_0=a\, (m_1,m_2,\ldots, m_d)$ (here $a$ is the lattice spacing). 
In contrast, if the walker starts from  
a nearest neighbour of the target, then the MFPT drecreases monotonically with increasing $r$, 
approaching a universal limiting value $1$ as $r\to \infty$, indicating that the optinal resetting rate in this 
case is infinity. These interesting results on a lattice are not captured by the continuum theory. We have also
performed high precision numerical simulations on hypercubic lattices up to
$50$ dimensions
and find excellent matching with our analytical predictions.
For targets close to the starting points, in the limit of high dimensions
or large resetting rates, $\langle T\rangle_r(\vec R_0)$ can also be obtained by a simple
pedagogical Markov chain formulation.

Finally, in the absence of a target, we have also computed
exactly the position distribution in the nonequilibrium stationary state that also displays interesting
regimes for large $r$ which are not captured by the continuum limit.

Our results thus demonstrate, within the context of a single particle diffusion, that there are very interesting 
lattice effects in the presence of a nonzero 
resetting rate $r$. Since the continuum theory requries scaling the resetting rate $r\to r a^2$ to zero as the 
lattice spacing $a\to 0$, it misses many effects, in particular for large resetting rate $r$, that are captured by 
the lattice computations. Since many models of stochastic resetting have been studied in continuous
space in the literature, it would be natural to study them on lattices to see if there are interesting
effects due to the lattice structure of the underlying space.

There are other directions in which this work can be extended. Here we focused on a single free particle
on a $d$-dimensional lattice. Recent studies have shown that if $N$ independent particles diffuse
in continuous space starting from a common initial position $\vec R_0$ and get simultaneously reset to $\vec R_0$,
the system is driven at long times into a nonequilibrium stationary state with dynamically emergent correlations
between particles~\cite{BLMS2023,BMS2023,BLMS2024}. Using techniques developed in this paper, it would be interesting to
compute the stationary state for this simultaneously resetting independent $N$-particle system that diffuse 
on a lattice.
One may expect new interesting lattice effects on the mutual all-to-all correlations between
the particles in the stationary state, in particular in the limit $r\to \infty$ which is not captured
by the continuum theory.

\begin{acknowledgments}

  SNM thanks J. L. Lebowitz for asking the question about lattice MFPT
  during a talk at the 
Rutgers webinar series and A. J. Guttmann for a useful correspondence.
SNM acknowledges support from ANR Grant No. ANR- 23- CE30-0020-01 
EDIPS and the Alexander von Humboldt foundation for the
Gay Lussac-Humboldt prize that allowed a visit to the Physics
department at Oldenburg 
University, Germany where this work was performed. The simulations
were performed at the 
HPC cluster ROSA, located at the University of Oldenburg (Germany) and
funded by the DFG through its Major Research Instrumentation Program
(INST 184/225-1 FUGG) and the Ministry of Science and Culture (MWK) of the
Lower Saxony State.

\end{acknowledgments}

\appendix

\section{The continuum limit of the lattice MFPT in Eq. (\ref{mfpt_final})}
\label{Cont_lim_App}

To take the continuum limit of Eq. (\ref{mfpt_final}), it is convenient to start from
the most general expression for MFPT in \eqref{mfpt_ltg.1} that reads
\begin{equation}
\langle T\rangle_r(\vec R_0)=\frac{1}{r}\left[ \frac{\tilde{P}_0(\vec 0, \vec 0, r)}{\tilde{P}_0(\vec 0,
\vec R_0, r)}-1\right]\, .
 \label{mfpt_ltg.A1}
\end{equation}
For $\tilde{P}_0(\vec 0,
\vec R_0, r)$, it is now convenient to use the integral representation
\eqref{int_rep.2} which reads
\begin{widetext}
\begin{equation}
\tilde{P}_0(\vec 0, \vec R_0, r)=\frac{1}{(r+2d)}\int_0^{\infty} dt\, e^{-t}\,
{\displaystyle \prod_{i=1}^d} \int_{-\pi}^{\pi} \frac{dk_i}{2\pi}\,
\exp\left[ \frac{2\,t}{(r+2d)}\, \cos(k_i)\right] \, e^{-i\, \vec k\cdot {\vec R_0}/a}\, .
\label{int_rep.a1}
\end{equation}
Next we make the change of variable $k_i= a\, q_i$. This gives
\begin{equation}
\tilde{P}_0(\vec 0, \vec R_0, r)=\frac{a^d}{(r+2d)}\int_0^{\infty} dt\, e^{-t}\,
{\displaystyle \prod_{i=1}^d} \int_{-\pi/a}^{\pi/a} \frac{dq_i}{2\pi}\,
\exp\left[ \frac{2\,t}{(r+2d)}\, \cos(a\, q_i)\right] \, e^{-i\, \vec q\cdot {\vec R_0}}\, .
\label{int_rep.a2}
\end{equation}
As $a\to 0$, we expand $\cos(a q_i)\approx 1- a^2 q_i^2/2$ to rewrite this as
\begin{equation}
\tilde{P}_0(\vec 0, \vec R_0, r)\approx \frac{a^d}{(r+2d)}\, \int_0^{\infty} dt\, e^{-rt/(r+2d)}\,
{\displaystyle \prod_{i=1}^d} \int_{-\pi/a}^{\pi/a} \frac{dq_i}{2\pi}\,
e^{- \frac{a^2 t}{(r+2d)}\, q_i^2} \, e^{-i\, \vec q\cdot {\vec R_0}}\, .
\label{int_rep.a3}
\end{equation}
Since $a\to 0$, the limits of the integrals can be pushed to $\mp \infty$ to leading order and
one can make use of the identity
\begin{equation}
{\displaystyle {\prod_{i=1}^d}} \int_{-\infty}^{\infty} \frac{dq_i}{2\pi}\, e^{- \sigma^2\, q_i^2/2- i \vec q\cdot
\vec R_0}= \frac{1}{(\sigma\, \sqrt{2\pi})^d}\, e^{-R_0^2/{2\sigma^2}}\, .
\label{iden_A1}
\end{equation}
This gives
\begin{equation}
\tilde{P}_0(\vec 0, \vec R_0, r)\approx \frac{(r+2d)^{d/2-1}}{(4\pi)^{d/2}}\, \int_0^{\infty} \frac{dt}{t^{d/2}}\, 
\exp\left[ - \frac{r}{r+2d}\, t - \frac{(r+2d)\, R_0^2}{4\, a^2\, t}\right]\, .
\label{P0_A1}
\end{equation}
\end{widetext}
We next perform the integral using the following identity~\cite{Grads}
\begin{equation}
\int_0^{\infty} x^{\nu-1} e^{-\beta/x- \gamma\, x}= 2\, \left(\frac{\beta}{\gamma}\right)^{\nu/2}\, 
K_{\nu}\left(\sqrt{4 \beta \gamma}\right)\, .
\label{Bessel_A1}
\end{equation}
Using this identity in \eqref{P0_A1} and setting $r= a^2\, \tilde{r}$ gives finally
\begin{equation}
\tilde{P}_0(\vec 0, \vec R_0, r)\approx \frac{R_0^\nu}{(2\pi)^{d/2}\, a^{2\nu}\, 
{\tilde{r}}^{\nu/2}}\, K_\nu\left(R_0 \, \sqrt{\tilde{r}}\right)
\label{P0_A2}
\end{equation}
where $\nu=1-\frac{d}{2}$.
This gives us the denominator in Eq. (\ref{mfpt_ltg.A1}). To obtain the numerator we just take the
limit $R_0\to 0$ in \eqref{P0_A2} by using the small argument asymptotics of the Bessel function 
in Eq. (\ref{BesselK_asymp}). Putting these results together in Eq. (\ref{mfpt_ltg.A1})
and rescaling $T= \tilde{T}/a^2$, we obtain our desired result in Eq. (\ref{result_cont.1}).

\end{document}